\documentclass[usenatbib]{aa}
\usepackage{graphicx}	
\usepackage{amsmath}	
\usepackage{hyperref}

\usepackage{txfonts}
\newcommand{\gaea}{{\sc GAEA2023~}}
\newcommand{\gaeax}{{\sc GAEA2020X~}}
\newcommand{\gaeaf}{{\sc GAEA2020F~}}
\begin{document}
\title{Tracing the Quenching Journey across Cosmic Time}
\author{Gabriella De Lucia\inst{1,2}\fnmsep\thanks{e-mail:gabriella.delucia@inaf.it},
  Fabio Fontanot\inst{1,2}, Lizhi Xie\inst{3}\and Michaela Hirschmann\inst{4,1}
}
\institute{INAF - Astronomical Observatory of Trieste, via G.B. Tiepolo 11, 
        I-34143 Trieste, Italy
        \and
        IFPU - Institute for Fundamental Physics of the Universe, via Beirut 2,
        34151, Trieste, Italy
        \and
        Tianjin Normal University, Binshuixidao 393, 300387, Tianjin, People’s
        Republic of China
        \and
        Institute for Physics, Laboratory for Galaxy Evolution, EPFL,
        Observatoire de Sauverny, Chemin Pegasi 51, 1290 Versoix, Switzerland }

   \date{Received ???, 2023; accepted ???, 2023}

   \abstract{We present the latest version of the GAlaxy Evolution and Assembly
     (GAEA) theoretical model of galaxy formation. Our new model now combines
     (i) an updated treatment of AGN feedback including an improved modelling
     of cold gas accretion on super-massive black holes and an explicit
     implementation of quasar winds; and (ii) a treatment for both cold and hot
     gas stripping from satellite galaxies. We show that our latest model
     version predicts specific star formation rate distributions that are in
     remarkable agreement with observational measurements in the local
     Universe. Our updated model predicts quenched fractions that are in very
     nice agreement with observational measurements up to $z\sim 3-4$, and a
     turn-over of the number densities of quenched galaxies at low stellar
     masses that is in qualitative agreement with current observational
     estimates. We show that the main reasons for the improved behaviour with
     respect to previous renditions of our model are the updated treatment for
     satellites at low galaxy masses ($<10^{10}\,{\rm M}_{\sun}$) and the
     inclusion of quasar winds at intermediate to large stellar masses
     ($>10^{10}\,{\rm M}_{\sun}$). However, we show that the better treatment
     of the star formation threshold, due to our explicit partitioning of the
     cold gas in its atomic and molecular components, also plays an important
     role in suppressing excessive residual star formation in massive galaxies.
     While our analysis is based on a selection of quiescent galaxies that
     takes advantage of the information about their star formation rate, we
     demonstrate that the impact of a different (colour-colour) selection is
     not significant up to $z \sim 3$, at least for galaxies above the
     completeness limits of current surveys. Our new model predicts number
     densities of massive quiescent galaxies at z > 3 that are the largest
     among recently published state-of-the-art models. Yet, our model
     predictions appear still to be below post-JWST observational
     measurements. We show that the expected cosmic variance is large, and can
     easily accommodate some of the most recent measurements.}


\titlerunning{Quenching across Cosmic Time}\authorrunning{G.~De Lucia et al.}
   
   \maketitle
%
\section{Introduction}
\label{sec:intro}

It is a well established result that a fraction of the galaxy population at all
cosmic epochs are not actively forming stars, and typically have red colours
and old stellar populations. This fraction is also known to vary with galaxy
environment and cosmic epoch. In the local Universe, where rather accurate
estimates of the star formation activity can be obtained using spectroscopic
information for large statistical samples of galaxies, these are shown to
follow a `bimodal' distribution, with galaxies being either blue and actively
forming stars or red and `quenched' \citep[e.g.][just to cite the milestone
  papers on the subject based on the Sloan Digital Sky
  Survey]{Blanton_etal_2003,Kauffmann_etal_2003,Baldry_etal_2004}. The
`transition' region, sometimes dubbed `green valley', is relatively empty and
its position does not vary significantly in different environments, suggesting
that if galaxies move from one region to another, this must occur on a
relatively short time-scale \citep{Wetzel_Tinker_Conroy_2012} and must be
driven by a process that is effective in different environments.

At earlier cosmic epochs, the separation between quiescent and active galaxies
cannot be made using direct measurements of the star formation activity (at
least not for large galaxy samples), and it is instead based on colour-colour
selections. This, together with errors associated with photometric redshifts,
make the separation between passive and active galaxies subject to larger
uncertainties than when using direct estimates of star formation rates
(SFRs). Deep photometric surveys have enabled studies of the luminosity, mass,
and colour distribution of galaxiex at higher redshift, confirming that colour
bimodality persists up to $z\sim 4$ and beyond
\citep[e.g.][]{Brammer_etal_2009,Whitaker_etal_2011,Muzzin_etal_2013,Weaver_etal_2023}.
The advent of the James Webb Space Telescope (JWST) has allowed us to push
these studies to even earlier cosmic epochs revealing a large population of
massive and apparently quiescent galaxies at $z>3$, in excess of previous
observational inferences \citep[e.g.][]{Carnall_etal_2023,Valentino_etal_2023}.

This rich amount of information has triggered numerous theoretical studies and
highlighted, over the years, important disagreements between observational
measurements and theoretical predictions. These have, in turn, lead to
significant improvements of published models. Since early renditions of
semi-analytic models, it was realized that an additional source of energy at
the centre of massive systems is required to avoid the formation of overly
massive and star forming galaxies. In early models
\citep[e.g.][]{Kauffmann_etal_1999,Benson_etal_2003,DeLucia_etal_2004}, this
was implemented in a very simplistic way by suppressing gas cooling above some
critical halo mass/circular velocity. Later theoretical developments, including
a physical treatment of the impact of feedback from Active Galactic Nuclei
(AGN), have confirmed that this process plays a crucial role in suppressing
the cooling flows at the centre of massive clusters, thereby reconciling the
massive end of the galaxy stellar mass function with observations and leading
to a sequence of red and massive galaxies that is qualitatively consistent with
data \citep[e.g.][and many
  others]{Croton_etal_2006,Bower_etal_2006,Sijacki_etal_2007,Lagos_etal_2008,Dubois_etal_2013,Puchwein_and_Springel_2013}.

Early detailed comparisons between theoretical predictions and colour
distributions of galaxies in the local Universe, split by centrals and
satellites, showed a significant excess of passive satellites with respect to
observational estimates \citep{Weinmann_etal_2006}. It was soon realized that
these discrepancies were shared by all state-of-the-art theoretical models of
galaxy formation published in those years \citep{Fontanot_etal_2009}. A revised
treatment of satellite galaxies lead to some improvements, but reproducing the
colour/SFR distributions observed for satellites and central galaxies in the
local Universe has long remained, and to some extent still represents, an
important challenge for both semi-analytic and hydro-dynamical simulations of
galaxy formation \citep[e.g.][and references
  therein]{Hirschmann_etal_2014,Wang_etal_2018,DeLucia_etal_2019,Xie_etal_2020,Donnari_etal_2021,Bravo_etal_2023}. The
challenges become even stronger when trying to reproduce the estimated
fractions of quiescent galaxies at earlier cosmic epochs
\citep{DeLucia_etal_2019,Donnari_etal_2021,Lustig_etal_2023,Weaver_etal_2023},
or when focusing on the cluster environment where the galaxy population is
dominated by satellites \citep{Bahe_etal_2017,Kukstas_etal_2023}.

In this paper, we present an updated version of our GAlaxy Evolution and
Assembly ({\sc GAEA}) model, that combines our recent improved treatments for
AGN feedback and ram-pressure stripping of hot and cold gas from satellite
galaxies, and an explicit treatment for partitioning the cold gas in its atomic
and molecular components. We make a critical assessment of the predicted trends
for passive galaxies, as a function of physical properties (stellar mass
primarily), redshift, and environment. In a companion paper
\citep{Xie_etal_2024}, we discuss the emergence and physical origin of the
first quiescent galaxies.

The layout of the paper is as follows: in Section~\ref{sec:simsam}, we present
the model and the simulations used in this paper. In Section~\ref{sec:z0}, we
compare predictions from our new model with observational estimates of the
distribution of specific star formation rate (sSFR) and passive fractions in
the local Universe, while in Section~\ref{sec:highz} we consider observational
estimates of passive fractions at earlier cosmic epochs. In
Section~\ref{sec:clust}, we present a comparison between our model predictions
and estimated passive fractions in galaxy clusters at $z\sim 1$. Finally, in
Section~\ref{sec:discconcl}, we discuss our results and give our conclusions.

\section{The simulation and the galaxy formation model}
\label{sec:simsam}

GAEA\footnote{Details about the model, and access to a selection of data
  products, can be found at: https://sites.google.com/inaf.it/gaea/} traces the
formation and evolution of galaxies in a cosmological framework, using
state-of-the-art prescriptions for the evolution of different baryonic
components, coupled to substructure based merger trees extracted from
high-resolution N-body simulations. The galaxy formation model builds on that
presented in the original paper by \citet{DeLucia_and_Blaizot_2007}, but it has
been updated significantly over the years. In particular, our model includes:
\begin{enumerate}
\item[(i)] a detailed treatment of the non-instantaneous recycling of gas,
  energy, and metals, with an accurate accounting of their timings depending on
  stellar lifetimes, which enables the tracing of individual metal abundances
  \citep{DeLucia_etal_2014};
\item[(ii)] a parametrization of stellar feedback that is partially based on
  results from hydro-dynamical simulations, and that we have shown allows us to
  reproduce the observed galaxy stellar mass function up to $z \sim 3$
  \citep{Hirschmann_etal_2016};
\item[(iii)] an explicit partition of the cold gas in its atomic and molecular
  components, tuned to reproduce the measured HI and H$_2$ galaxy mass
  functions in the local Universe \citep{Xie_etal_2017};
\item[(iv)] a careful tracing of the angular momentum exchanges between
  different components, and a treatment for the non instantaneous stripping of
  the cold and hot gaseous components associated with galaxies infalling onto
  larger systems \citep{Xie_etal_2020};
\item[(v)] an improved model for cold gas accretion on super massive black
  holes (BHs) and AGN driven outflows that, in the framework of our model,
  reproduces the measured evolution of the AGN luminosity function and the
  so-called AGN downsizing trend \citep{Fontanot_etal_2020}.
\end{enumerate}

The last two implementations enumerated above represent so far two independent
branches of our model, that we have now merged into the model presented in this
work. Our model also features the possibility to adopt a variable stellar
Initial Mass Function \citep[IMF - ][]{Fontanot_etal_2017b,Fontanot_etal_2024},
that we do not consider in this work where we only present results based on a
Chabrier IMF.

\begin{table*}
  \caption{Values of the relevant parameters used in \citet{Xie_etal_2020} and
    \citet{Fontanot_etal_2020}, and corresponding values adopted in the model
    considered in this study. For each parameter, we also briefly explain the
    meaning and give the appropriate reference.}
  \label{tab:params}

  \renewcommand{\footnoterule}{}
  \centering
  \begin{tabular}{llccccc}
    \hline
    & meaning & reference & \gaeax & \gaeaf & \gaea \\
    \hline
    \\
    \multicolumn{2}{l}{stellar feedback parameters as defined in \citet{Hirschmann_etal_2016}}  \\
    \hline
    $\epsilon_{\rm reheat}$ & reheating efficiency & Eq.~14        & 0.3  & 0.13  & 0.28  \\
    $\epsilon_{\rm eject}$ & ejection efficiency & Eq.~15         & 0.1  & 0.23  & 0.10  \\
    $\gamma_{\rm reinc}$   & re-incorporation efficiency & Eq.~2       & 1.0  & 0.68  & 0.99 \\
    \hline
    \\
    \multicolumn{2}{l}{other parameters as defined in \citet{Xie_etal_2020}}  \\
    \hline
    rps$_{\rm time}$ & timescale of hot gas ram-pressure stripping & Eq.~9         & 444 & --  & 400  \\
    kesi$_{\rm slow}$   & ratio between specific angular momentum of & & & & \\
                      & gas cooling through `slow mode' and that of the halo & Sect.~2.1       & 1.4  & 1.0  & 1.19 \\
    kesi$_{\rm rapid}$   & ratio between specific angular momentum of & & & & \\
                      & gas cooling through `rapid mode' and that of the halo & Sect.~2.1       & 3.0  & 1.0  & 1. 38\\
    \hline
    \\
    \multicolumn{2}{l}{BH accretion parameters as defined in \citet{Fontanot_etal_2020}} \\    
    \hline
    $\kappa_{\rm radio}/10^{-5}$  & hot gas black hole accretion efficiency  & Eq.~2  & 1.0  & 0.6  & 1.36   \\
    $f_{\rm lowJ}/10^{-3}$  & cold gas angular momentum loss efficiency & Eq.~4         & -- & 6. & 3.18 \\
    $f_{\rm BH}/10^{-3}$   & black hole accretion rate from reservoir & Eq.~12          & -- & 0.09  & 0.11 \\
    $\epsilon_{\rm qw}/100.$ & quasar wind efficiency & Eq.~15      & -- & 3.2 & 4.86 \\ 
    $f_{\rm cen}/10^{-3} $    & fraction of ISM added to the BH reservoir  &  Sec. 2.4  & -- & 3.0  & 3.39 \\
                           & as a consequence of AGN-driven outflows &               &    &      & \\
    \hline
  \end{tabular}
  
\end{table*}

In previous work, we have shown that our reference model \citep[][and later
  versions]{Hirschmann_etal_2016} is able to reproduce a large number of
important observational constraints. These include the galaxy stellar mass
function up to $z\sim 7$ and the cosmic star formation rate density up to
$z\sim 10$ \citep{Fontanot_etal_2017}, the relation between galaxy stellar mass
and gas metallicity and its secondary dependence as a function of star
formation rate and gas mass \citep{DeLucia_etal_2020}, and the observed
evolution of the galaxy mass - gas/star metallicity relations
\citep{Hirschmann_etal_2016,Fontanot_etal_2021}. Our model is of course not
without problems. In previous work, we have emphasized in particular the
difficulty to reproduce the bimodality in colour/specific star formation rate
observed in the local Universe \citep{Hirschmann_etal_2016,Fontanot_etal_2020},
and the tendency of the model to underpredict the observed fractions of passive
galaxies at higher redshift \citep{DeLucia_etal_2019}. As mentioned in
Section~\ref{sec:intro}, these difficulties are shared to different extent by
virtually all recently published theoretical models of galaxy formation. We
will show, in the following, that these problems are largely overcome by
  the latest rendition of GAEA presented in this work.

We refer to the original papers cited above (and references therein) for a
detailed description of the specific implementations adopted.  As for AGN
feedback, we use what in \citet{Fontanot_etal_2020} is referred to as the {\sc
  F06-GAEA} set of parametrizations, that include: (i) a modelling for the
inflow of cold gas towards the central massive black holes driven by star
formation in the central regions of galaxies; (ii) an accretion rate that is
determined by the viscous accretion time-scale; (iii) an empirical scaling
relation between the mass loading of quasar winds and the bolometric
luminosity. In Table~\ref{tab:params}, we list the values of the relevant
parameters adopted in \citet[][\gaeax hereafter]{Xie_etal_2020} and
\citet[][{\sc GAEA2020F}]{Fontanot_etal_2020}, and the corresponding values
adopted in the new model whose results are discussed in this work ({\sc
  GAEA2023} in the following). All parameters that are not included in the
table have been left unchanged with respect to the model versions published in
our previous work. The primary observables that have been considered to
recalibrate our model are: the observed galaxy stellar mass function and its
evolution up to $z\sim 3$, the HI and H$_2$ mass functions in the local
Universe, the evolution of the AGN luminosity function up to $z\sim 4$. When
calibrating our model parameters, we also make sure that the mass-metallicity
relations predicted at $z=0$ have approximately the correct normalization. We
will come back to the impact of these parameter modifications on our model
predictions in the final discussion section. In Appendix~A, we show the main
figures used to calibrate our model parameters.

The results presented in this paper are based on the Millennium Simulation
\citep{Springel_etal_2005}. This simulation follows 2,160$^3$ dark matter
particles in a box of 500~${\rm Mpc}\,{\rm h}^{-1}$ on a side, and assumes
cosmological parameters consistent with WMAP1 ($\Omega_\Lambda=0.75$,
$\Omega_m=0.25$, $\Omega_b=0.045$, $n=1$, $\sigma_8=0.9$, and $H_0=73 \, {\rm
  km\,s^{-1}\,Mpc^{-1}}$). While more recent measurements provide slightly
different cosmological parameters (in particular, a larger value for $\Omega_m$
and a lower value for $\sigma_8$), we do not expect these differences to affect
significantly our model predictions, once the model parameters have been
retuned to reproduce a specific set of observational results \citep[see
  also][]{Wang_etal_2008,Guo_etal_2013}. In a forthcoming paper, we will
re-address this issue presenting results from the same model considered here
but applied to a simulation with cosmological parameters consistent with a
Planck cosmology. To verify convergence of our physical model, we will also use
in this work the MillenniumII Simulation \citep{Boylan-Kolchin_etal_2009}. This
assumes the same cosmological model and uses the same number of particles of
the Millennium Simulation, but has a smaller volume (one-fifth the size of the
Millennium, i.e. 100~Mpc~${\rm h}^{-1}$), and 125 times better mass resolution
(the particle mass is $6.9\times10^6\,{\rm M_{\sun}}\,{\rm h}^{-1}$).

\section{The specific star formation rate distribution in the local Universe}
\label{sec:z0}

We start, in this section, by considering model predictions at z=0. In
particular, we focus here on the distribution of sSFR and on the fraction of
passive galaxies, as a function of the galaxy stellar mass. The observational
data shown in this section are based on data from the Sloan Digital Sky Survey
(SDSS) DR8, cross-correlated with the JHU-MPA DR7
catalogue\footnote{http://www.mpa-garching.mpg.de/SDSS/DR7/} and with the DR7
version of the group catalogue by \citet{Yang_etal_2007}. Stellar masses
included in the JHU-MPA DR7 catalogue are estimated based on fits to the
photometry following the philosophy of \citet{Kauffmann_etal_2003} and
\citet{Salim_etal_2007}. SFRs are based on \citet{Brinchmann_etal_2004}.
Aperture corrections, that are computed by fitting the photometry of the outer
regions of galaxies as detailed in \citet{Salim_etal_2007}, have been used to
convert to total SFR estimates.

\begin{figure*}
\centering
\resizebox{18cm}{!}{\includegraphics{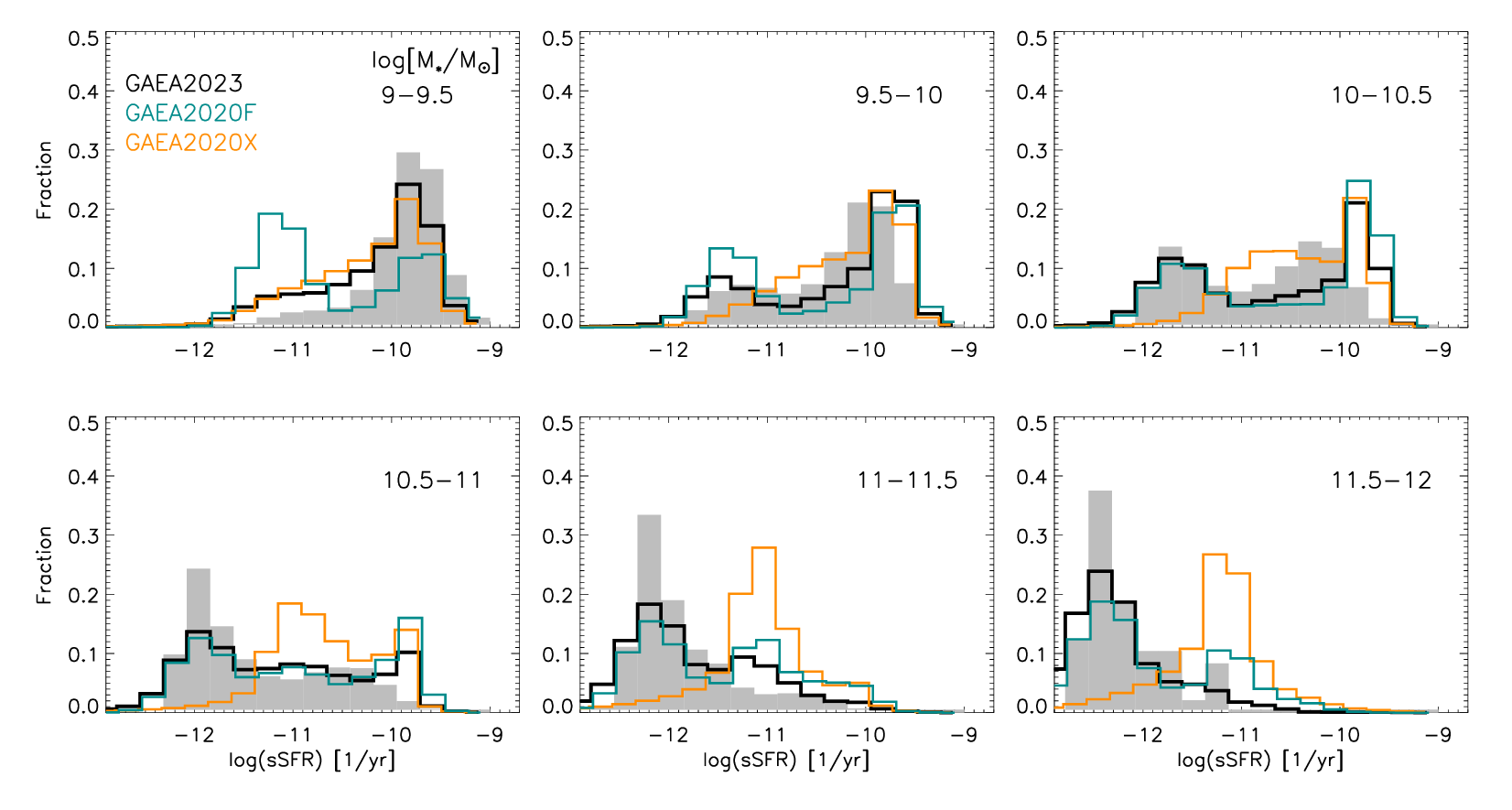}} 
\caption{Specific star formation rate distribution as predicted by the models
  indicated in the legend, and compared to observational estimates based on
  SDSS DR8 (gray shaded histograms - see text for details). Each panel
  corresponds to a different bin in galaxy stellar mass, as indicated in the
  top-right legend. The histograms corresponding to the \gaeax and \gaeaf
  models have been slightly shifted (by $0.02$~dex) to the left and right with
  respect to the \gaea model for clarity.}
\label{fig:ssfrdistr}
\end{figure*}

Fig.~\ref{fig:ssfrdistr} shows the sSFR distributions for bins of galaxy
stellar mass (different panels). Gray histograms show the observational
estimates, while open histograms correspond to predictions from different
versions of GAEA, as indicated in the legend. For the observational sample, we
have considered only galaxies in the redshift range $0.025 < z < 0.05$, where
we expect the sample to be approximately volume complete down to a galaxy
stellar mass $\sim 10^9\,{\rm M}_{\sun}$. Following \citet{Fontanot_etal_2020},
we have assigned an `upper observable value' for the SFR to each model galaxy
that has a theoretical SFR $< 10^{-4}\,{\rm M}_{\sun}\,{\rm yr}^{-1}$. For
these galaxies, we have adopted the following relation, that reproduces the
locus of the upper limits of passive galaxies in the SDSS sample:
\begin{displaymath}
\log[{\rm SFR}] = 0.5 \times \log[{\rm M}_{\rm star}] - 6.59.  
\end{displaymath}
The above relation has also been perturbed by assuming a lognormal scatter of
0.25 dex.

The figure shows that, as discussed in previous work, both the \gaeax and
\gaeaf versions of our model have problems in reproducing the observed sSFR
distributions. In particular, {\sc GAEA2020X} follows relatively well the
observational distribution for the lowest galaxy mass bin considered, although
the predicted distribution is somewhat skewed towards low values of sSFR with
respect to data. However, this is the range where incompleteness in the
observed sample and/or our treatment of upper observable values of SFR become
relevant. At intermediate masses, the \gaeax model does not reproduce well the
observed bimodality in sSFR, and for the most massive galaxies the predicted
distributions are shifted towards sSFR values larger than observed. We had
highlighted similar problems in \citet{Hirschmann_etal_2016}, where we also
noted that we were unable to solve the excess of star formation in massive
galaxies with simple modifications of the (radio mode) AGN feedback. The reason
for this, as discussed in \citet{Fontanot_etal_2020}, is that this star
formation excess is not related to late gas cooling nor to gas brought in by
satellites accretion. We found that the main reason for this excess residual
activity was the large gaseous content of the main progenitors of these
galaxies at $z\sim2$. Fig.~\ref{fig:ssfrdistr} shows that the inclusion of AGN
driven winds reduces significantly, albeit not removing completely, the excess
of activity in massive galaxies. This feedback process also introduces a clear
bimodality in the distributions of galaxies at intermediate stellar
masses. However, for the lowest mass bins considered, the \gaeaf model
over-predicts significantly the number of galaxies with low sSFR values.

The \gaea model presented in this paper captures well the observed trends, for
all galaxy mass bins considered. At the lowest stellar masses, model
predictions are very close to those from the \gaeax model, with no excess of
passive galaxies. At intermediate masses, there is a clear bimodality, with a
minimum at log[sSFR]$\sim -11\,{\rm yr}^{-1}$, consistent with observations. No
significant excess of star formation activity is found for the most massive
galaxies. Considering that we are not making any attempt to mimick in detail
the observational selection, and taking into account uncertainties in
observational estimates of SFRs and aperture corrections, the level of
agreement shown in Fig.~\ref{fig:ssfrdistr} between observational data and our
\gaea model is remarkable.

The improved agreement with data of the \gaea model with respect to previous
model renditions can be explained as follows: for low-mass galaxies, the better
regulation of star formation is driven by the improved treatment of
environmental processes introduced in the \gaeax model. In fact, as noted
above, predictions from the latter model are very similar to those from \gaea
in this mass range. In contrast, satellite galaxies tend to be over-quenched in
the \gaeaf model due to the instantaneous stripping of their hot gas reservoir
at the time of accretion, combined with a rather efficient supernovae
feedback. The slight modification of parameters between the \gaeax and \gaea
models also leads to slightly larger gas fractions (and therefore larger values
of star formation) for low-mass galaxies, resulting in a slight improvement
with respect to the \gaeax model.

At the most massive end, the inclusion of an efficient feedback channels
associated with quasar winds triggered by mergers and disk instability, leads
to a sSFR distribution that is peaked at values comparable to those observed
but leaves a non negligible excess of galaxies with sSFR~$\sim 10^{-11}\,{\rm
  yr}^{-1}$ in the \gaeaf model. We find that this excess is removed in the
\gaea model thanks to an the overall better improvement of the density
threshold for star formation. In fact, the prescriptions adopted in \gaeaf lead
to a very bursty behaviour of the star formation, with relatively large amounts
of gas accumulating for some time-steps and then leading to intense short
bursts of star formation. The partition of the cold gas in its atomic and
molecular gas components, and the assumption that star formation can only take
place in regions corresponding to a certain surface density of molecular
hydrogen (for details, see \citealt{Xie_etal_2017}), limits significantly this
behaviour.

At intermediate galaxy stellar masses, the more pronounced bimodality with
respect to the \gaeax model, as well as the good agreement between the position
of the predicted and observed `passive' peak of the sSFR distribution can also
be largely ascribed to efficient ejections of gas due to quasar winds.

\begin{figure*}
\centering
\resizebox{18cm}{!}{\includegraphics{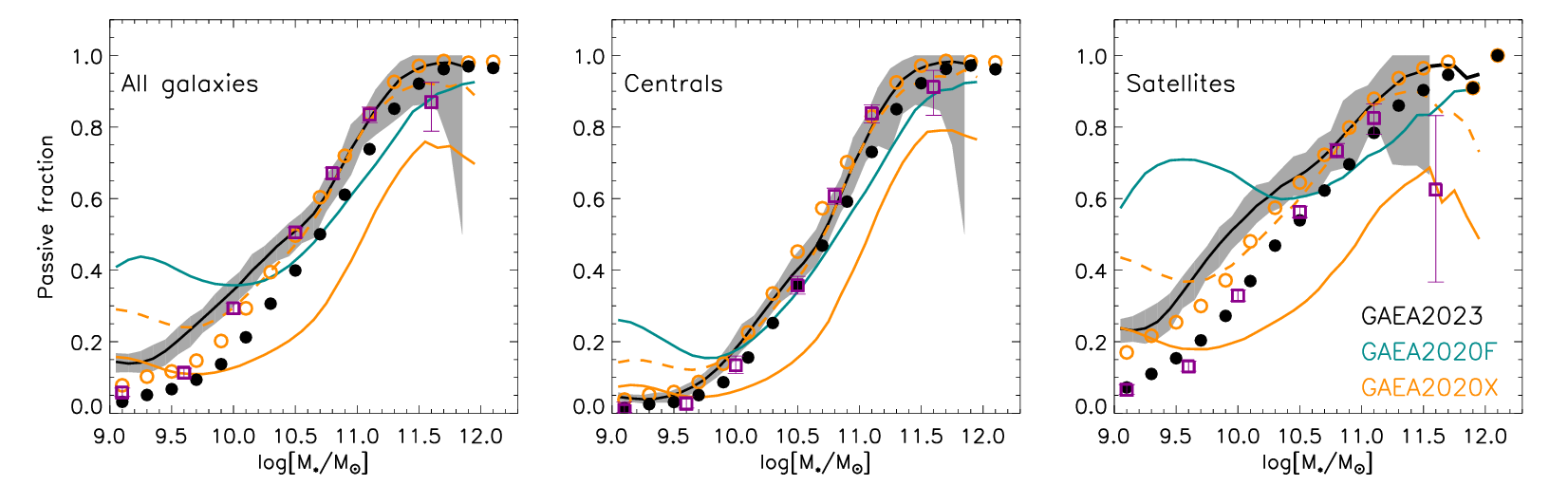}} 
\caption{Passive fractions for all galaxies (left panel), central galaxies
  (middle panel), and satellites (right panels). Symbols correspond to
  observational measurements. In particular: squares correspond to estimates
  from \citet{Davies_etal_2019} based on the GAMA survey, and adopt a log(sSFR)
  threshold of $-10.5\,{\rm yr}^{-1}$ to identify passive fractions. Filled and
  empty circles correspond to measurements based on SDSS DR8 and are obtained
  assuming a threshold of $-11\,{\rm yr}^{-1}$ and $-10.6\,{\rm yr}^{-1}$,
  respectively. Solid lines correspond to predictions from different models, as
  indicated in the legend, assuming a log(sSFR) threshold of $-11.0\,{\rm
    yr}^{-1}$ to select passive galaxies. The gray regions show the
  minimum-maximum fractions obtained when considering predictions from the
  \gaea model from 125 independent boxes, each of $100$~Mpc~${\rm h}^{-1}$ on a
  side. The dashed line shows predictions from the {\sc GAEA2020X} model
  corresponding to a log(sSFR) threshold of $-10.6\,{\rm yr}^{-1}$, as adopted
  in \citet{Xie_etal_2020}.}
\label{fig:fquenchz0}
\end{figure*}

Fig.~\ref{fig:fquenchz0} shows the fraction of passive galaxies for all
galaxies (left panel), central galaxies only (middle panel), and satellites
(right panel). Solid lines show predictions from the different models, as
indicated in the legend, and all correspond to a sSFR threshold of
$10^{-11}\,{\rm yr}^{-1}$ to identify passive galaxies (as noted above, this
describes well the separation between the two sSFR peaks for the observational
data). The dashed line corresponds to the \gaeax model and to a threshold of
$10^{-10.6}\,{\rm yr}^{-1}$. This is the threshold that was adopted in
\citet{Xie_etal_2020} and better describes the separation between the two sSFR
peaks in that model. Symbols correspond to observational estimates: squares
with error bars are based on GAMA \citep{Davies_etal_2019} and assume a sSFR
threshold equivalent to $10^{-10.5}\,{\rm yr}^{-1}$ to select passive galaxies;
filled and open circles show estimates based on SDSS data assuming a threshold
of $10^{-11}\,{\rm yr}^{-1}$ and $10^{-10.6}\,{\rm yr}^{-1}$, respectively.
The figure shows that the latest version of our model reproduces very nicely
the observational estimates in the local Universe. In particular: for central
galaxies, the predicted passive fractions decrease monotonically with
decreasing galaxy stellar mass, down to masses $\sim 10^9\,{\rm M}_{\sun}$. The
same trend is found for satellite galaxies, also in quite good agreement with
observational estimates. Previous versions of our model over-predict the
passive fraction of satellites below galaxy stellar masses $\sim
10^{9.5}-10^{10.5}\,{\rm M}_{\sun}$, and under-predict the fraction of passive
massive satellites. This is more significant for the {\sc GAEA2020X} model,
especially when considering the same sSFR threshold in all models. To give an
idea of the cosmic variance for our model, we divide the simulated volume of
the Millennium in 125 independent box of 100~Mpc~${\rm h}^{-1}$ on a side, and
compute the expected quiescent fractions in each independent sub-volume. The
gray regions shown in each panel represent the regions between the minimum and
maximum quiescent fractions obtained. The variance is relatively small at these
scales, but not negligible, and it becomes significant at the largest galaxy
stellar masses (the rarest systems).

\section{Passive galaxies beyond z=0}
\label{sec:highz}

In this section, we present model predictions for passive fractions beyond the
local Universe, and compare them with different observational estimates. We
emphasize that these are {\it true predictions}, because these data have not
been considered when tuning the model parameters. We use a sSFR selection for
model galaxies. Specifically, we select as passive galaxies all those with sSFR
smaller than $0.3\times t^{-1}_{\rm Hubble}$ \citep{Franx_etal_2008}, which
corresponds to $\sim 10^{-11}\,{\rm yr}^{-1}$ at $z=0$. In
Section~\ref{sec:discconcl}, we will discuss the impact of a different
selection.

\begin{figure}
\centering
\resizebox{8cm}{!}{\includegraphics{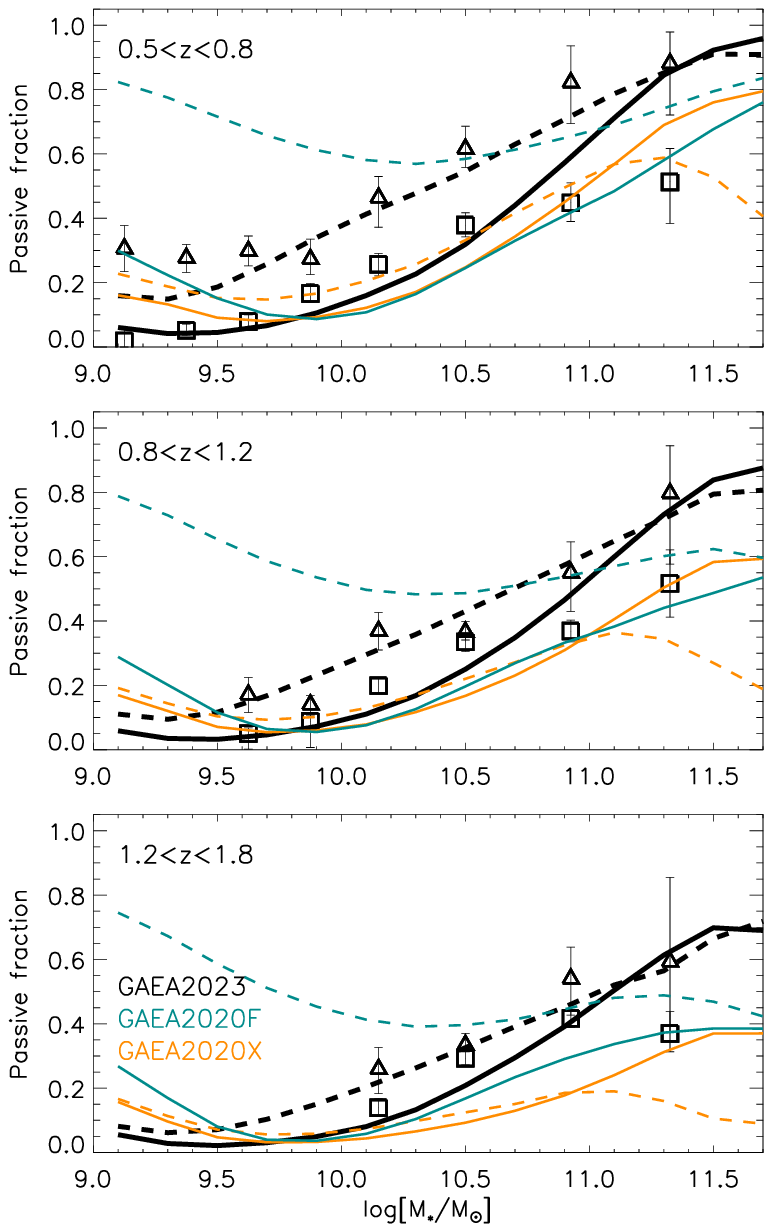}} 
\caption{Fraction of passive centrals (solid lines) and satellites (dashed
  lines) as predicted by the models considered in this study. Symbols with
  error bars (squares are for centrals and triangles for satellites) show
  observational estimates by \citet{Fossati_etal_2017}, based on five
  CANDELS/3D-HST fields.} 
\label{fig:fossati}
\end{figure}

Fig.~\ref{fig:fossati} shows the predicted passive fractions in three different
redshift bins, for central (solid lines) and satellite galaxies (dashed
lines). Symbols with error bars show the corresponding observational estimates
by \citet{Fossati_etal_2017}. These have been computed using spectroscopic and
grism redshifts in the five CANDELS/3D-HST fields, and multiwavelength
photometry to classify galaxies as passive. The central/satellite information
has been obtained by assigning, to each galaxy, probability functions computed
using a mock galaxy catalogue that reproduces the 3D-HST sample selection and
redshift accuracy. The fraction of passive galaxies tends to increase, both in
the models and in the data, with decreasing redshift. For the \gaeax model, the
fractions of passive centrals and satellites are very similar at all redshifts
and for the entire galaxy mass range considered, but for the most massive
galaxies where the fraction of passive satellites turns over (as it does at
$z=0$ - see right panel of Fig.~\ref{fig:fquenchz0}). This trend is in contrast
with observational estimates that always find a fraction of quenched satellites
larger than that of quenched centrals, with the difference decreasing with
increasing redshift. In addition, the overall predicted fractions of passive
satellites from the \gaeax model are lower than estimated, at all
redshifts. This is true also for the fraction of passive centrals at the
highest redshifts considered. The fractions of passive centrals predicted by
the \gaeaf model are very similar to those predicted by the {\sc GAEA2020X}
model, while the fractions of passive satellites are larger. This brings model
predictions in better agreement with observational estimates for the most
massive satellites, but also leads to a significant over-prediction of passive
satellites with masses below $10^{10}\,{\rm M}_{\sun}$, over the entire
redshift range considered. Our latest model version exhibits the best agreement
with the observational data considered in Fig.~\ref{fig:fossati}, with a
similar decrease of the difference between satellites and centrals with
increasing redshift. The predicted fractions of passive centrals tend to be
larger than observed for the most massive galaxies, but the uncertainties here
are larger due to small number statistics and larger uncertainties in the
classification of centrals and satellites.

\begin{figure*}
\centering
\resizebox{18.5cm}{!}{\includegraphics{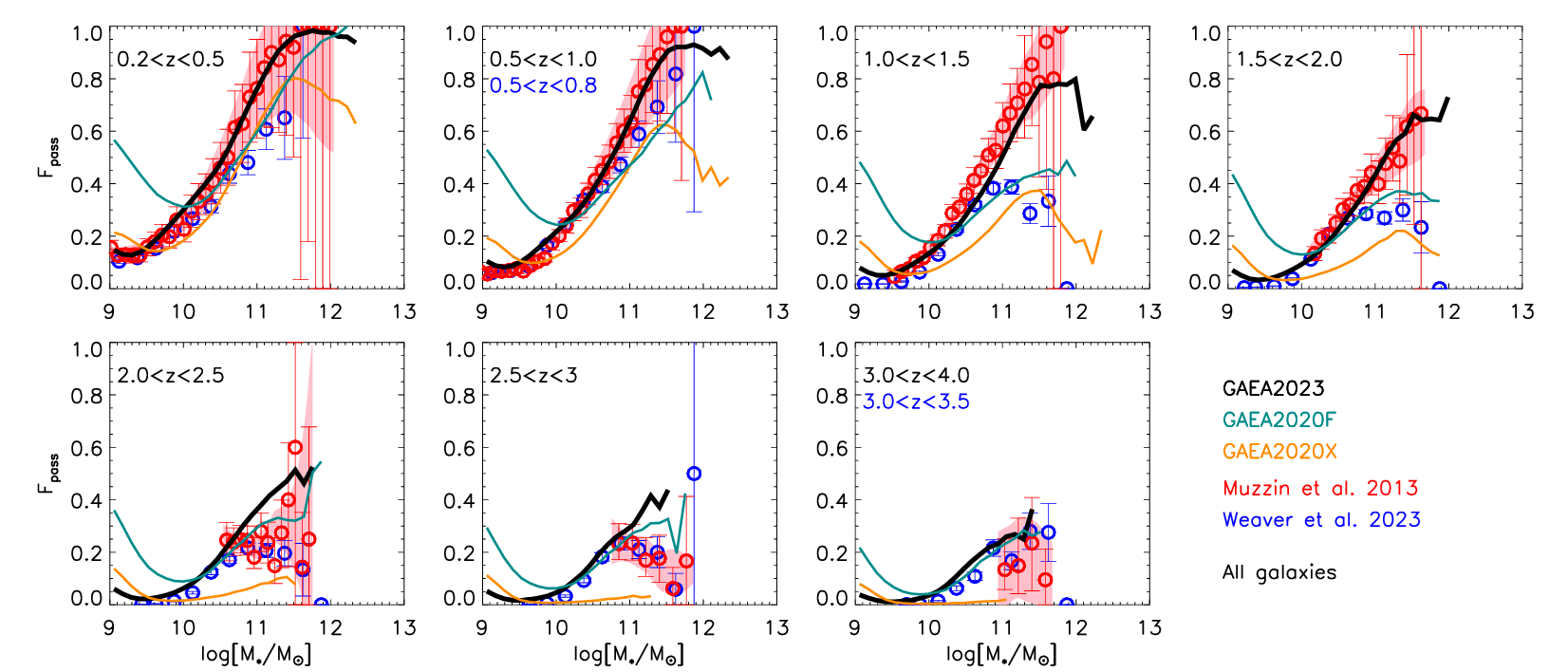}} 
\caption{Passive fraction at different cosmic epochs. Lines correspond to
  predictions from the models considered in this study. The shaded area and red
  symbols with error bars show observational estimates from
  \citet{Muzzin_etal_2013}. Specifically, the shaded areas show the $1/{\rm
    V}_{\rm max}$ measurements, while the open symbols with error bars show the
  maximum-likelihood measurements (see bottom panels of Fig. 6 in
  \citealt{Muzzin_etal_2013}). The blue symbols with error bars correspond to
  the observational estimates by \citet{Weaver_etal_2023}. A sSFR cut has been
  employed to select model passive galaxies (see text for details), while the
  observational samples are based on different colour-colour selections, a SED
  fitting estimate of the galaxy stellar mass, and photometric redshifts
  estimates.}
\label{fig:muzzinfrac}
\end{figure*}

Pushing the comparison to earlier cosmic epochs, we consider the observational
estimates by \citet{Muzzin_etal_2013} and by \citet{Weaver_etal_2023}. The
former\footnote{Available at:
  http://cosmos.phy.tufts.edu/$\sim$danilo/MuzzinEtal2013} are based on a
sample of $\sim 100,000~K_s$ selected galaxies in the COSMOS/UltraVISTA field,
with photometric redshifts estimated using photometry covering the wavelength
range $0.15-8.0\mu$m. The mean estimated uncertainty of photometric redshifts
is $\delta z/(1+z) = 0.013$, and the estimated fraction of catastrophic
outliers is only $\sim 1.6$~per cent. Stellar masses have been estimated using
SED fitting and a Kroupa IMF, consistent with the one adopted in our
model. Passive galaxies have been identified using a ($UVJ$) colour-colour
selection. Estimates\footnote{Available at: https://zenodo.org/records/8377094}
by \citet{Weaver_etal_2023} are based on the most recent release of the COSMOS
catalogue \citep[the COSMOS2020, ][]{Weaver_etal_2022}. The sample, including
approximately $1$~million galaxies, is complete down to $10^9\,{\rm M}_{\sun}$
out to $z\sim 3$, and is supported by extensive photometry ranging from the UV
to $8\,\mu$m and including deep near-infrared UltraVISTA imaging and Subaru
Suprime-Cam intermediate bands. The quoted photo-z uncertainties are
$\delta z/(1+z) < 0.01$ at $i\sim 20$, and $<0.04$ at $i\sim 26$ AB.
Passive galaxies have been selected adopting a ($NUVrJ$) colour-colour
selection, and galaxy stellar masses have been estimated assuming a Chabrier
stellar IMF.

Fig.~\ref{fig:muzzinfrac} shows a comparison between our model predictions and
these observational estimates. The red open symbols with error bars and shaded
regions show the maximum likelihood measurements and the $1/{\rm V}_{\rm max}$
measurements by \citet{Muzzin_etal_2013}. The blue symbols with error bars show
the observational estimates by \citet{Weaver_etal_2023}. Our model predictions
are shown by lines of different colours, as indicated in the legend. For this
comparison, we have selected all model galaxies from snapshots falling in the
redshift bins used by Muzzin et al. These match well those considered in Weaver
et al. but for the second and last redshift bins shown (see figure legend - the
redshift bins adopted by \citealt{Weaver_etal_2023} are written in
blue). However, we have verified that the comparison is not affected
significantly by this slight mismatch.

The \gaeax model exhibits a progressive worsening of the underprediction of
passive fractions with increasing redshift. Only about 20 per cent or less of
the most massive galaxies are predicted to be passive at $1.5 <z< 2.0$, against
a fraction that is estimated to be as large as $\sim 60$~per cent by
\citet{Muzzin_etal_2013}. Less that 1 per cent of the galaxies in the entire
volume considered from the \gaeax run are predicted to be passive at
$z>2.5$. The inclusion of an explicit feedback mode associated with AGN (in the
form of quasar driven winds), mainly relevant for radiatively efficient BH
accretion, increases significantly the fraction of passive galaxies in the
\gaeaf model. For galaxies more massive than $\sim 10^{10}\,{\rm M}_{\sun}$,
the passive fractions predicted by this model are lower than observational
estimates by \citet{Muzzin_etal_2013} but in quite nice agreement with
estimates by \citet{Weaver_etal_2023}. For less massive galaxies, this model
over-predicts significantly the observational estimates.  As shown in
Fig.~\ref{fig:fossati}, the overprediction of passive fractions for lower mass
galaxies is due primarily to an excess of quenched satellites in this
model. Predictions from our new model, \gaea, are in very good agreement with
the observational estimates by \citet{Muzzin_etal_2013}, over the entire
redshift range considered, but over-predict the passive fractions measured by
\citet{Weaver_etal_2023} at galaxy stellar masses larger than $\sim
10^{11}\,{\rm M}_{\sun}$ at $1<z<2$. It is interesting that the observational
measurements considered differ significantly over this redshift and mass
range. It is difficult to identify one single possible reason for this
difference: the two studies are based on a different colour-colour selection
($UVJ$ in the case of Muzzin et al. and $NUVrJ$ in the case of Weaver et al.),
but there are other important differences both in the data used and in the data
analysis. E.g. the deeper NIR selection employed by \citet{Weaver_etal_2023}
might be more sensitive to dusty star-forming galaxies. In fact, many massive
systems that are identified in COSMOS2020 are red systems best-fit by dusty
star forming spectral templates (Weaver J., private communication). While the
differences between these published estimates can be considered as a rough
indication of the systematic uncertainties for quenched fractions at large
stellar masses, we note that measurements by Weaver et al. show a rather strong
evolution between $z\sim 1$ and $z\sim 1.5$. This appears to be inconsistent
with stellar population studies of massive quiescent galaxies in the local
Universe finding very old stellar ages (and therefore early quenching
times). At low galaxy stellar masses, the observational estimates considered
are consistent, and our {\sc GAEA2023} model is in very nice agreement with the
observed trends down to the lowest masses shown in the figure, with some hint
for a turnover of passive fractions at the lowest masses accessible through the
Millennium Simulation. We will come back to this point in
Section~\ref{sec:discconcl}.

\begin{figure*}
\centering
\resizebox{18cm}{!}{\includegraphics{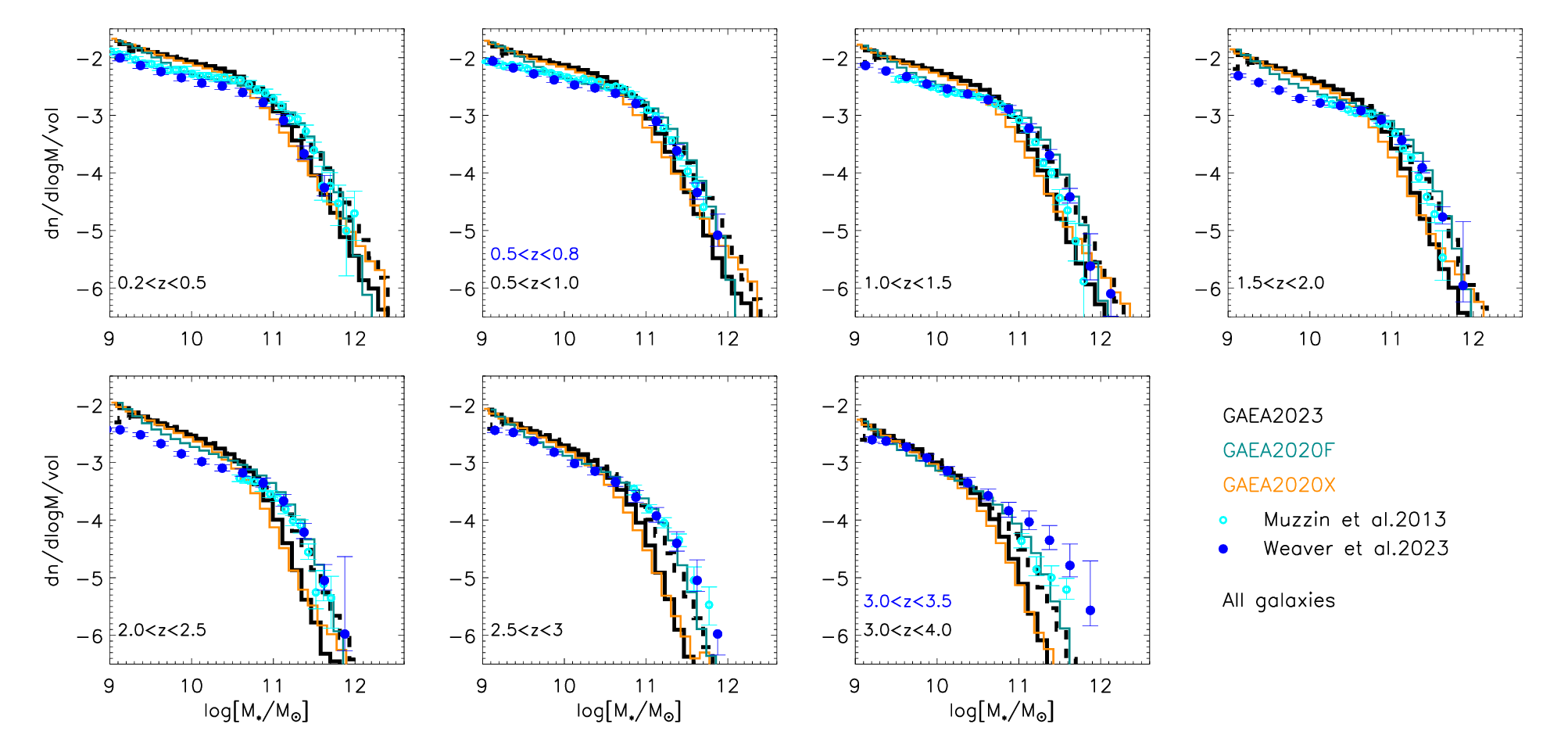}} 
\caption{Galaxy stellar mass function at different cosmic epochs. Symbols with
  error bars correspond to the observational estimates published in \citet[][in
    this case we are showing only the $1/{\rm V}_{\rm max}$ estimates published
    in this study]{Muzzin_etal_2013} and \citet{Weaver_etal_2023}. Solid lines
  correspond to predictions from the models considered in this study, as
  indicated in the legend. The dashed line in each panel shows, for the {\sc
    GAEA2023} model only, the corresponding predictions obtained assuming that
  the stellar masses have uncertainties that are Gaussian distributed with a
  width of $0.25$~dex.}
\label{fig:muzzingmfall}
\end{figure*}

\begin{figure*}
\centering
\resizebox{18cm}{!}{\includegraphics{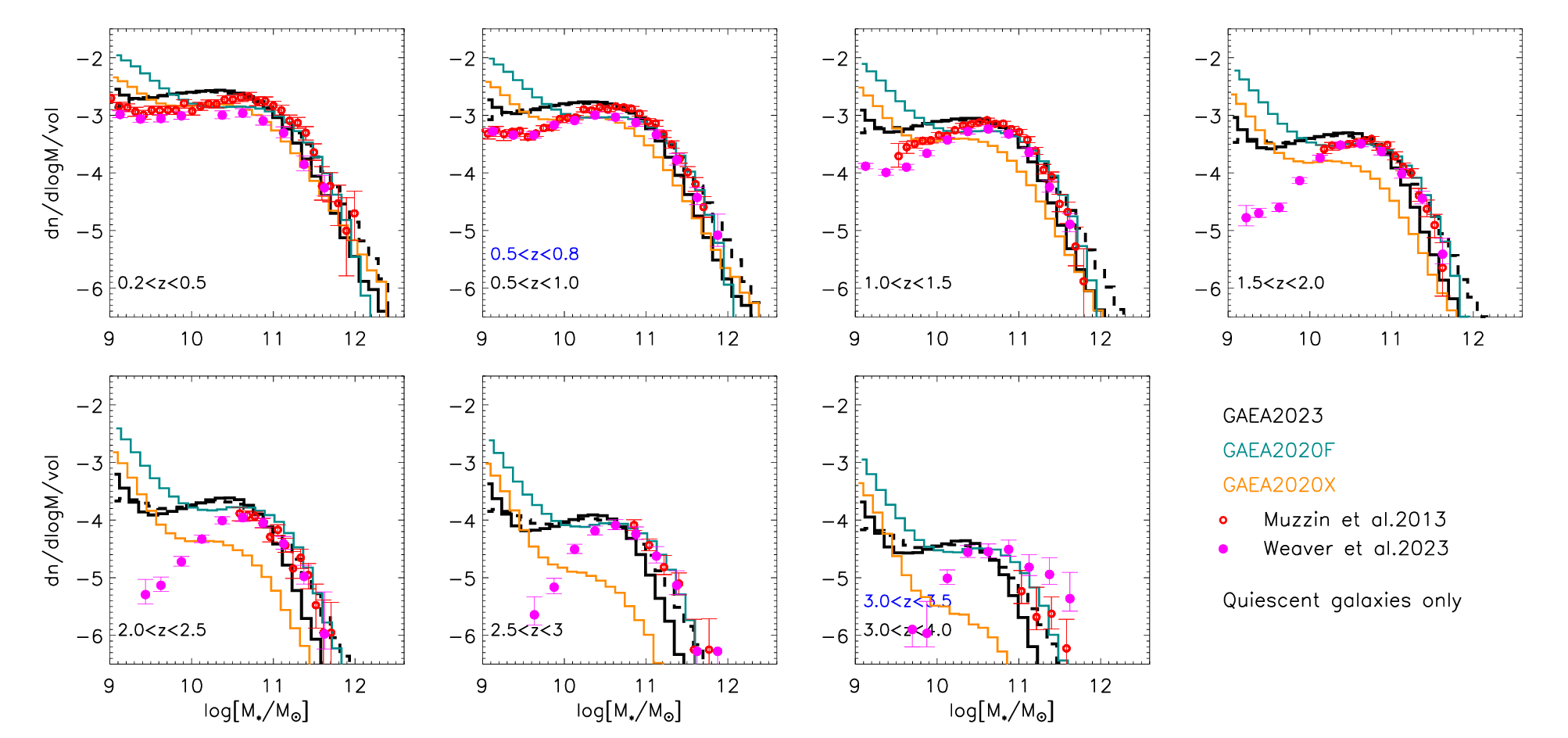}} 
\caption{As in Fig.~\ref{fig:muzzingmfall}, but for quiescent galaxies only.}
\label{fig:muzzingmfq}
\end{figure*}

In addition to the quenched fractions shown above, it is instructive to
consider the corresponding mass distributions from which they have been
computed, i.e. the galaxy stellar mass functions for all and for quiescent
galaxies. A comparison between our model predictions and the galaxy mass
functions published in \citet{Muzzin_etal_2013} and in \citet{Weaver_etal_2023}
is shown in Figs.~\ref{fig:muzzingmfall} and
\ref{fig:muzzingmfq}. Observational estimates are shown by symbols with error
bars, while lines correspond to our model predictions as indicated in the
legend. For the \gaea model only, we show both the intrinsic model predictions,
as well as the corresponding predictions obtained assuming that stellar masses
have uncertainties that are Gaussian distributed with a width of $0.25$~dex
(dashed lines). We note that, while this is a typical {\it statistical}
uncertainty quoted, it is well known that stellar mass estimates based on SED
fitting depend on a number of assumptions including metallicity, stellar
population models, stellar initial mass function, and star formation history
\citep[see e.g.][]{Conroy_etal_2009}, and that these can result into larger
uncertainties and even systematic trends as a function of galaxy stellar mass
\citep{Mitchell_etal_2013}. As emphasized above, we have used observational
estimates of the galaxy stellar mass function, up to $z\sim 3$, to constrain
our model parameters for the \gaea and \gaeax models. Therefore, the relatively
good agreement that is shown in Fig.~\ref{fig:muzzingmfall}, when considering
the galaxy mass function for all galaxies predicted by these models, does not
come as a surprise. The slight excess of galaxies below the knee of the mass
function, that is visible in particular in the top panels, is already visible
in our previous papers (see e.g. Fig.~A1 in \citealt{Xie_etal_2020}), but
becomes somewhat less evident when considering a compilation of independent
observational estimates (see Appendix A). The deficit of massive galaxies that
is evident for these two models, particularly in the bottom panels, is more
susceptible to the statistical errors associated with the estimates of stellar
mass, as shown by the dashed line in each panel. It is interesting to note that
the \gaeaf model, that has been tuned primarily to reproduce the AGN luminosity
function rather than the galaxy stellar mass function, predicts both lower
number densities of galaxies below the knee and larger number densities for the
most massive galaxies. At the highest redshift considered, our \gaea model
appears to be in quite good agreement with observational estimates by
\citet{Muzzin_etal_2013} while underpredicting the number densities of the most
massive galaxies when compared to observational measurements by
\citet{Weaver_etal_2023}. In their work, Weaver et al. comment about the larger
number densities of massive galaxies at these redshifts, when compared to
earlier studies, and speculate that these might be due to the presence of
proto-clusters in the field.

As for the galaxy mass function of passive galaxies
(Fig.~\ref{fig:muzzingmfq}), both the \gaeax and the \gaea models seem to work
reasonably well at the lowest redshift bin considered, with the \gaeax model
slightly underpredicting the number densities of quiescent galaxies around the
knee and showing a turn-over below stellar masses $\sim 10^{9.5}\,{\rm
  M}_{\sun}$. At the same redshift, the \gaeaf model exhibits a clear excess of
passive galaxies below $\sim 10^{10}\,{\rm M}_{\sun}$. These trends become
stronger at higher redshifts, with a dramatic steepening of the galaxy mass
function of quiescent galaxies below $\sim 10^{10}\,{\rm M}_{\sun}$ for the
\gaeaf model. As we have discussed earlier, this is mainly driven by an excess
of passive satellite galaxies. A turn over of the number densities of quiescent
galaxies is predicted also by the \gaea and \gaeax models, but it is moved
towards lower masses and does not appear to be excluded by available
data. Deeper observations at intermediate to high redshift can provide
important constraints on these models. In Section~\ref{sec:discconcl}, we will
discuss if and how the turn over at low galaxy stellar masses is affected by
the resolution of the adopted simulation.  As reflected in the low passive
fractions presented above, the \gaeax model significantly underpredicts the
number densities of quiescent galaxies at high redshift, with the
under-prediction becoming more severe with increasing redshift. The latest
version of our model predicts number densities of passive galaxies that are in
rather good agreement with observational estimates up to $z\sim 3$, over the
entire range of masses sampled by the data considered in this figure, and no
significant under-prediction of the most massive quiescent galaxies when one
accounts for the uncertainties in the stellar mass estimates. In fact, as noted
above, the typical uncertainties can also be much larger than the fiducial
$0.25$~dex assumed for the dashed lines shown. At the highest redshift shown,
the number densities of massive quiescent galaxies predicted by our model,
appear to under-predict the observational measurements by
\citet{Weaver_etal_2023}. As mentioned above, these authors argue that the
larger number densities estimated at this redshift (with respect to other
literature estimates) might be due to proto-clusters in the field (in one case,
they also have spectroscopic confirmation). We will get back to a more detailed
comparison with observational measurements of massive quiescent fractions at
$z>3$ later on.

\section{Passive galaxies in clusters at $z\sim 1$}
\label{sec:clust}

As mentioned in Section~\ref{sec:intro}, state-of-the-art models struggle to
reproduce in particular the estimated fraction of quiescent galaxies in
overdense regions, where the galaxy population is expected to be dominated by
satellite galaxies \citep[][see also Fig.~2 in
  \citealt{Xie_etal_2020}]{Bahe_etal_2017,Kukstas_etal_2023}. In this section,
we discuss how our model predictions compare to observational estimates based
on a subset of clusters from the GOGREEN survey
\citep{Balogh_etal_2017,Balogh_etal_2021}. This is based on homogeneous deep
imaging and spectroscopy of 21 galaxy groups and clusters at $1 < z <
1.5$. Below, we restrict the observational sample to the ten massive clusters
at $1.5 <z < 1.0$ that were analysed in \citet{vanderBurg_etal_2020}. Cluster
masses, determined from Jeans modelling as described in detail in
\citet{Biviano_etal_2021}, range between $10^{14.1}$ and $10^{14.8}\,{\rm
  M}_{\odot}$. In the Millennium Simulation, there are $\sim 300$ to $\sim 700$
haloes with ${\rm M}_{\rm 200c}$\footnote{This is the mass within a radius
  corresponding to an overdensity 200 times the critical density of the
  Universe.} in the same mass and redshift range of the GOGREEN cluster
considered. \citet{Kukstas_etal_2023} recently compared the same observational
measurements with three state-of-the-art suites of hydrodynamical simulations
(BAHAMAS+MACSIS, EAGLE+Hydrangea, IllustrisTNG) showing that simulations
generally reproduce well the stellar mass function of quenched galaxies in the
field, but all struggle to reproduce the corresponding measurements in the
cluster environment. In particular, all simulations considered in the work by
\citet{Kukstas_etal_2023} over-predict the fraction of quenched low-mass
satellite galaxies and two out of the three suites considered under-predict the
number densities of quenched massive galaxies, probably due to inefficient AGN
feedback.

To compare our model predictions with GOGREEN observational estimates, we have
selected all simulated haloes with mass larger than $10^{14}\,{\rm M}_{\odot}$
from the simulated volume, at all snapshots covering the redshift range of the
observed sample. Then, we have selected 25 sub-samples that match both the halo
mass and redshift distributions of the ten clusters used in
\citet{vanderBurg_etal_2020}. As in this observational work, we also consider
as cluster members only galaxies that are within a physical distance of $1$~Mpc
from the brightest cluster galaxies.

\begin{figure}
\centering
\resizebox{8cm}{!}{\includegraphics{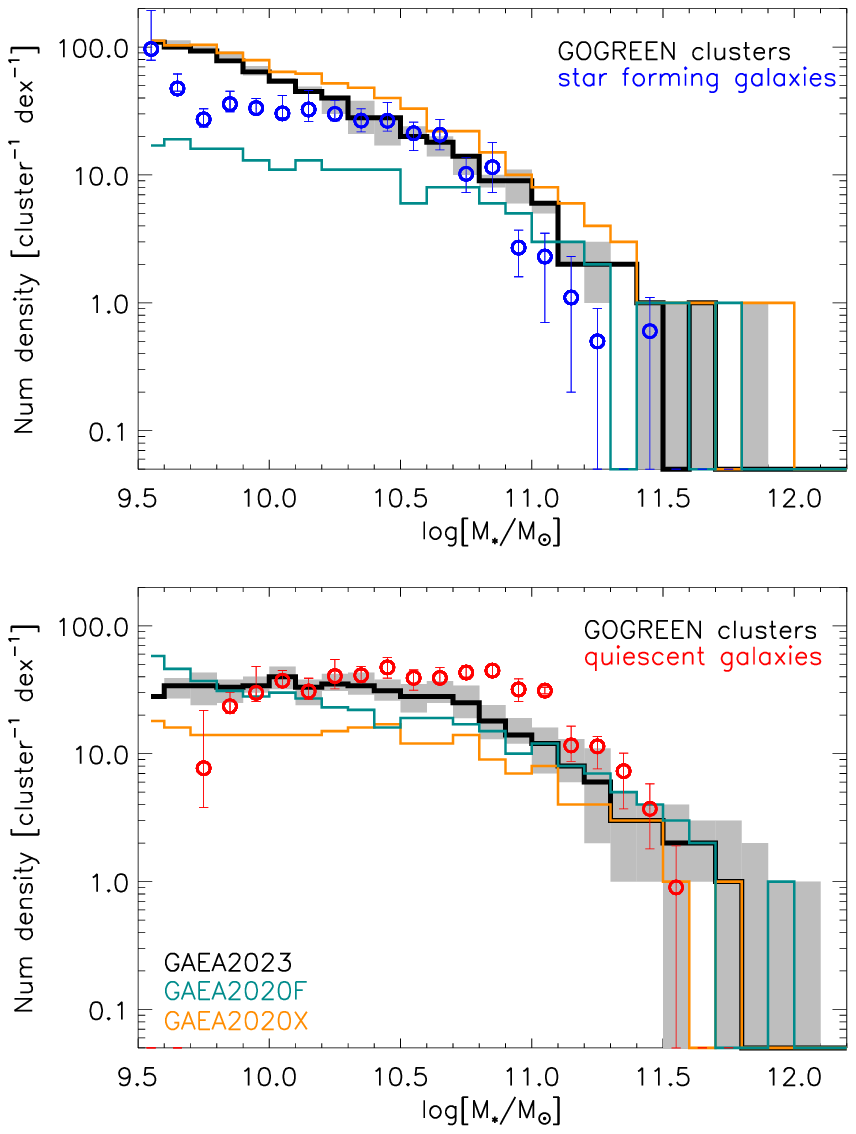}} 
\caption{Galaxy stellar mass function for the star forming (top panel) and
  quiescent galaxies (bottom panel) in clusters at $1 < z < 1.4$. Solid lines
  show predictions from the models considered in this study, as indicated in
  the legend. Symbols with error bars show corresponding observational
  estimates published in \citet{vanderBurg_etal_2020}. The shaded regions show
  the area enclosed by the 10th and 90th percentiles of the distributions of
  number densities obtained, for each galaxy stellar mass bin, when considering
  25 samples of simulated haloes that match the mass and redshift distributions
  of the GOGREEN sample (see text for details). All model lines shown in this
  figure assume an uncertainty in stellar mass of $0.25$~dex.}
\label{fig:gogreengmf}
\end{figure}

Fig.~\ref{fig:gogreengmf} shows a comparison between the observational
estimates of the galaxy stellar mass function presented in
\citet{vanderBurg_etal_2020} and our model predictions. Symbols with error bars
show the galaxy mass function for star forming (upper panel) and quiescent
galaxies (lower panel), classified on the basis of their location in a $UVJ$
colour-colour diagram. Solid lines in each panel show the corresponding
predictions from our different models, as indicated in the legend. As in
previous sections, we have used a time-dependent sSFR cut to select quiescent
model galaxies. The shaded regions show the area enclosed between the 10th and
90th percentiles of the distribution of number densities obtained in the \gaea
model, for each galaxy stellar mass bin considered, when considering the 25
simulated samples discussed above. All model predictions shown in this section
assume stellar mass uncertainties of $0.25$~dex.

The figure clearly shows that previous versions of our model fail to reproduce
simultaneously the observational measurements for quiescent and for star
forming galaxies: the \gaeax model undepredicts the number densities of
quiescent galaxies at all stellar masses, particularly below the knee. The
predicted mass function for star forming galaxies is closer to observational
data, with a slight excess of galaxies where the data exhibit a `dip' at low
stellar masses, and an excess of more massive star forming galaxies. The {\sc
  GAEA2020F} model undepredicts the number densities of quiescent galaxies
around the knee, and underpredicts significantly the number densities of star
forming galaxies below the knee. The \gaea model exhibits the best agreement
with data, especially when considering the effect of sample variance. However,
it still undepredicts slightly the number densities of quiescent galaxies around
the knee, and overpredicts slightly the number densities of star forming galaxies
at low-stellar masses.

\begin{figure}
\centering
\resizebox{8cm}{!}{\includegraphics{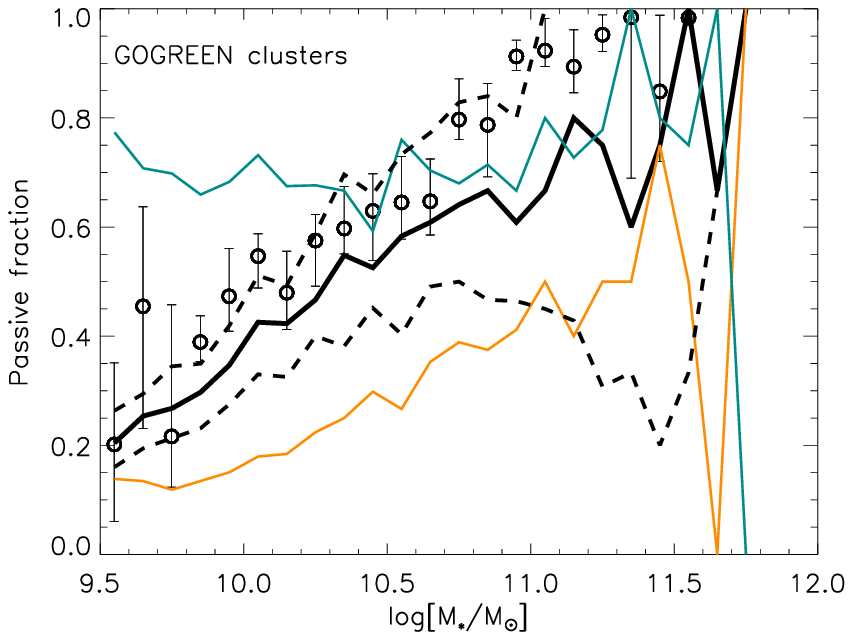}} 
\caption{Passive fraction in the cluster environment at $z\sim1$. Lines show
  predictions from the models considered in this study, while symbols with
  error bars show the observational estimates from
  \citet{vanderBurg_etal_2020}. The dashed lines show the area enclosed by the
  20th and 80th percentiles of the quiescent fractions obtained, for each
  galaxy stellar mass bin, when considering 25 samples of simulated haloes that
  match the mass and redshift distributions of the GOGREEN sample (see
  text for details). We show the dashed lines only for the \gaea model for
  clarity. All model lines shown in this figure assume an uncertainty in
  stellar mass of $0.25$~dex.}
\label{fig:qfracgogreen}
\end{figure}

Fig.~\ref{fig:qfracgogreen} shows the passive fractions corresponding to the
mass functions considered above. The dashed lines (these are shown only for the
\gaea model for clarity - the dispersion predicted by the other model versions
considered is similarly large) represent the 20th and 80th percentiles of the
distributions obtained considering our 25 simulated samples. Consistently with
what found in the field, the \gaeax model under-predicts the fractions of
passive galaxies for the entire galaxy stellar mass range considered, with only
few samples predicting a fraction of massive passive galaxies consistent with
the data. The \gaeaf model predicts a quiescent fraction that is rather
constant as a function of galaxy stellar mass, consistent with data for the
most massive galaxies, and significantly larger than observed for low-mass
galaxies. The passive fractions predicted by the \gaea model are slightly below
the observational estimates when considering the average results, but
consistent with the data within the expected sample variance. Considering that
the selection of quiescent galaxies in the models and in the data is based on
different criteria, the agreement of our \gaea with data is overall remarkable.

\section{Discussion}
\label{sec:discconcl}

In this paper, we have presented an updated version of our GAEA theoretical
model for galaxy formation that combines two major implementations we have
published in the last years: (i) an improved treatment for satellite galaxies
that features a non instantaneous stripping of the cold and hot gas components
associated with galaxies being accreted onto larger systems, and a careful
tracing of the angular momentum exchanges between different baryonic components
\citep{Xie_etal_2020}; (ii) an improved modelling of cold gas accretion on
super-massive black holes and an explicit implementation of quasar winds
\citep{Fontanot_etal_2020}. We have shown that the combination of these
implementations leads to a remarkable agreement between our model predictions
and the observed distributions of sSFR in the local Universe (see our
Fig.~\ref{fig:ssfrdistr}). This represents a significant improvement over
previous versions of our model that were either predicting an excess of
low-mass quiescent galaxies (the \gaeaf model), or a non negligible offset
towards larger sSFR values for massive galaxies and the lack of a clear
bimodality at intermediate galaxy masses (the \gaeax model).

As explained in Section~\ref{sec:z0}, the main reasons for the improvement with
respect to our previous model renditions are the following: at low galaxy
stellar masses, the updated treatment adopted for satellite galaxies (as
implemented in \citealt{Xie_etal_2020}) reduces drastically the significant
excess of quenched galaxies found in the \gaeaf model.  At intermediate to
large galaxy stellar masses, the improvements are mainly driven by the
inclusion of AGN feedback in the form of quasar winds, with a non negligible
role played by an overall better treatment of the density threshold for star
formation.

\subsection{Impact of different selections for passive galaxies}

Throughout this work, we have used a classification of passive galaxies based
on the sSFR. While this is appropriate for the local Universe, the
observational selection of quiescent galaxies at $z>0$ is typically based on a
colour-colour selection. In particular, most previous work relies on the
rest-frame $U-V$ versus $V-J$ diagram
\citep[e.g.][]{Williams_etal_2009,Whitaker_etal_2011,Muzzin_etal_2013}. This
preference for a photometric selection is driven by the fact that rest-frame
colours can be easily calculated for any galaxy sample, while an accurate
estimate of the sSFR has stronger requirements on the depth and wavelength
coverage of the data. Given that the division into two populations is motivated
by the exhistence of a clear bimodality in physical/observable properties of
galaxies, the expectation is that a different selection for model galaxies does
not have a strong impact on the results discussed above. However, it is useful
and interesting to quantify explicitly the impact of a different galaxy
selection on the results discussed in the previous Sections.

\begin{figure*}
\centering \resizebox{18cm}{!}{\includegraphics{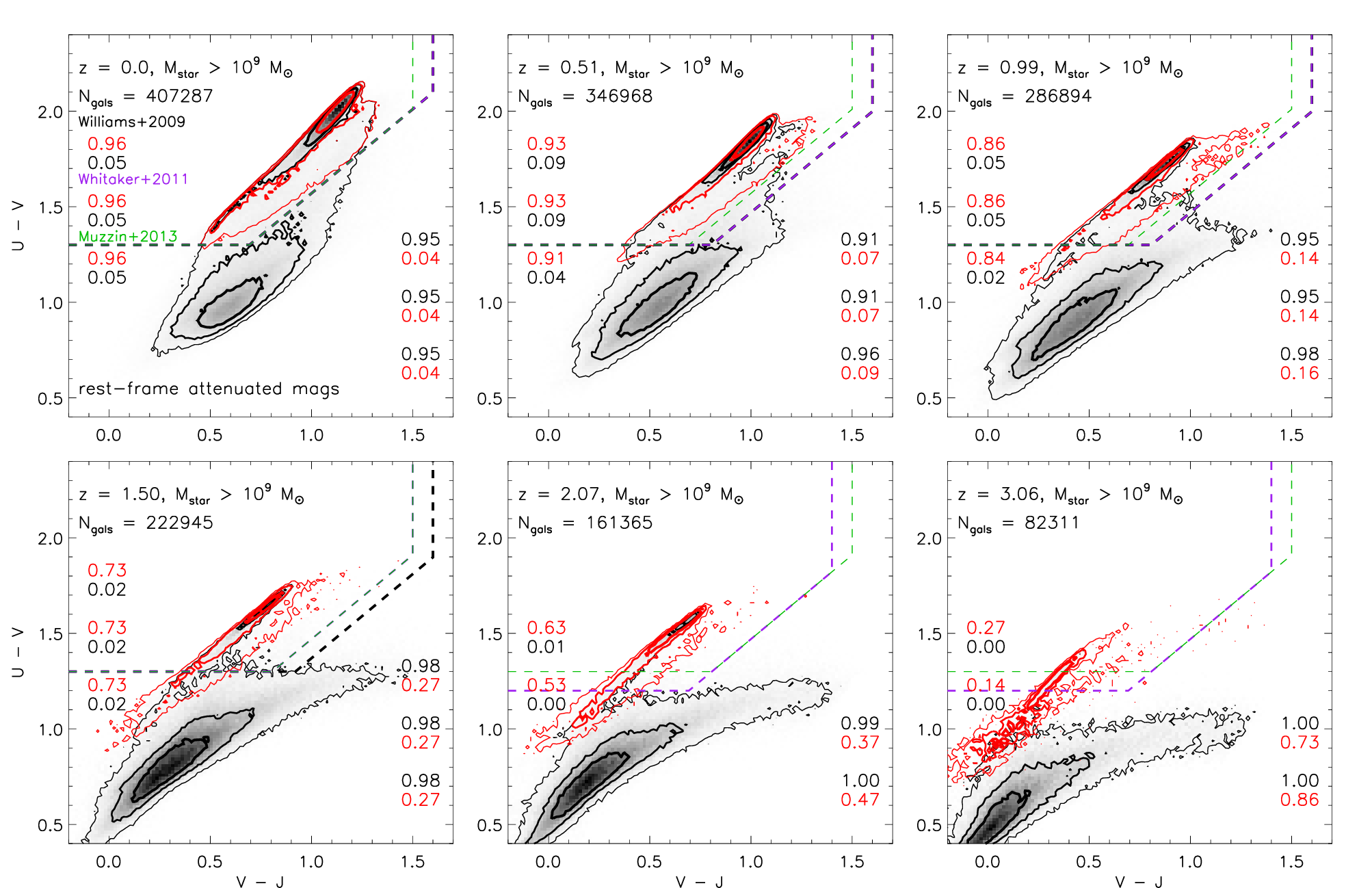}}
\caption{$U-V$ versus $V-J$ diagram for model galaxies at different cosmic
  epochs. Gray shaded regions show the distributions of all model galaxies more
  massive than $\sim 10^{9}\,{\rm M}_{\sun}$ at each redshift. Contours mark
  the regions enclosing 90, 60, and 30 per cent of the galaxies. The red
  contours show the corresponding contour levels for galaxies that are
  classified as quiescent according to their sSFR. Dashed lines indicate the
  regions typically considered for a photometric selection of passive galaxies,
  in different studies. The fractions given in each panel indicate the {\it
    completeness} and {\it contamination} of the passive and active samples
  selected using the $UVJ$ diagram (see text for details).}
\label{fig:uvj}
\end{figure*}

Fig.~\ref{fig:uvj} shows, in each panel, the distribution of all model galaxies
(gray shaded regions) at the redshift indicated in the legend. The dashed lines
indicate the box regions of the $UVJ$ diagrams that are typically used to
distinguish between star forming and quiescent galaxies. The black contours
show the regions enclosing 90, 60, and 30 per cent of the model galaxies
(increasing thickness of the lines). Red contours show the regions occupied by
galaxies that are classified as passive on the basis of their sSFR, as used in
the previous sections. For this analysis, we have only used a (representative)
sub-volume of the Millennium Simulation (about 10 per cent). In each panel, we
also give the total number of model galaxies used and the following
information:
\begin{itemize}
  \item on the left of each panel, in red, the {\it completeness}: the
    fractions of quiescent galaxies (on the basis of their sSFR) that fall in
    the passive region of the diagram. We also give the {\it contamination} in
    black, i.e. the fraction of star-forming galaxies that fall in the same
    passive region of the diagram;
  \item on the right in each panel, we give the same quantities but for the
    active region of the diagram.
\end{itemize}

\begin{figure*}
\centering \resizebox{18cm}{!}{\includegraphics{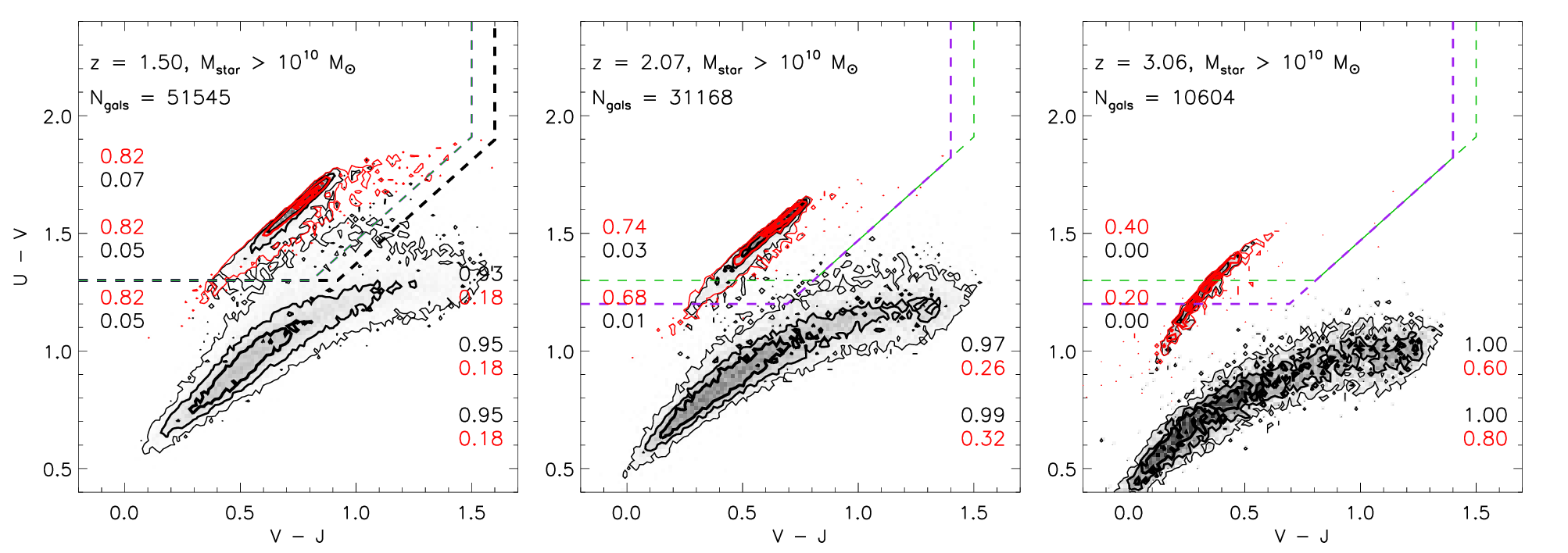}}
\caption{As in Fig.~\ref{fig:uvj} but only considering model galaxies more
  massive than $\sim 10^{10}\,{\rm M}_{\sun}$ and for the three highest
  redshifts considered.}
\label{fig:uvj10}
\end{figure*}

We give three different estimates of completeness and contamination associated
with different colour cuts widely adopted in the literature. Although slightly
different selections have been proposed in the literature, these do not lead to
significant differences in the completeness and contamination that we can
estimate using our model. Some differences are found only at the two highest
redshifts considered, where the selection suggested by
\citet{Whitaker_etal_2011} corresponds to systematically larger completeness
fraction with respect to the selection used in \citet{Muzzin_etal_2013}, with
no increase in contamination. The completeness of the passive sample selected
using the $UVJ$ diagram is very high, up to $z\sim 1.5$, decreasing to $\sim
60$ per cent at $z\sim 2$ and to only $\sim 20$ per cent at $z\sim 3$. We note
that these numbers are based on a sample of model galaxies including all
galaxies down to $\sim 10^{9}\,{\rm M}_{\sun}$, which is typically (much) below
the completeness limit of the observational data at intermediate-to-high
redshift. We show in Fig.~\ref{fig:uvj10} the same $U-V$ versus $V-J$ diagram,
but this time considering only model galaxies with galaxy stellar mass larger
than $\sim 10^{10}\,{\rm M}_{\sun}$, and for the three highest redshifts
considered. When considering only galaxies more massive than $\sim
10^{10}\,{\rm M}_{\sun}$, the completeness of the $UVJ$ classification goes up
to $\sim 70$ per cent at $z\sim 2$ and $20-40$ per cent at $z\sim 3$ depending
on the specific colour cuts adopted. We stress that, at these redshifts,
current observational samples are typically complete at even larger galaxy
stellar masses.

Considered the numbers given above, and the fact that the colour cuts applied
in different observational studies are often slightly modified to account
e.g. for slight zero-point differences of the available photometry, we can
conclude that a different selection of quiescent galaxies would not have a
significant impact on the agreement between model predictions and observational
estimates discussed in previous Sections up to $z\sim 2$. At higher redshifts,
the $UVJ$ selection fails to identify a significant fraction of galaxies that
have very low sSFR values (see also Fig. 1 in \citealt{Xie_etal_2024}). While
previous studies have demonstrated that this classical colour-colour selection
is indeed incomplete at $z>3$, the quoted `failure rate' is smaller than what
found for our model. We note that several model ingredients/assumptions might
have an impact on synthetic colours, e.g. assumptions about the stellar initial
mass function, stellar evolutionary tracks, and dust.

In future work, we plan to carry out a more detailed comparison with
observational data by mimicking the same colour-colour selections adopted in
observational studies, extending the analysis to alternative selections that
have been recently employed at high-z.

\subsection{Is there a turn-over of the number densities of low-mass quiescent galaxies?}

In Section~\ref{sec:highz}, we have shown that all models considered in this
study predict a turn-over of the quiescent galaxy stellar mass function at low
masses (see our Fig.~\ref{fig:muzzingmfq}). While this appears to be in
agreement with recent observational findings
\citep{Santini_etal_2022,Weaver_etal_2023}, the value at which this turnover
occurs for our model is very close to the resolution limit of the Millennium
Simulation. In order to quantify the impact of resolution, we have run our
\gaea model on the MillenniumII Simulation \citep{Boylan-Kolchin_etal_2009}. As
mentioned above, this simulation adopts the same cosmological model of the
Millennium but has 125 times better mass resolution, allowing to resolve
galaxies down to stellar masses $\sim 10^8\,{\rm M}_{\sun}$.

\begin{figure*}
\centering
\resizebox{18cm}{!}{\includegraphics{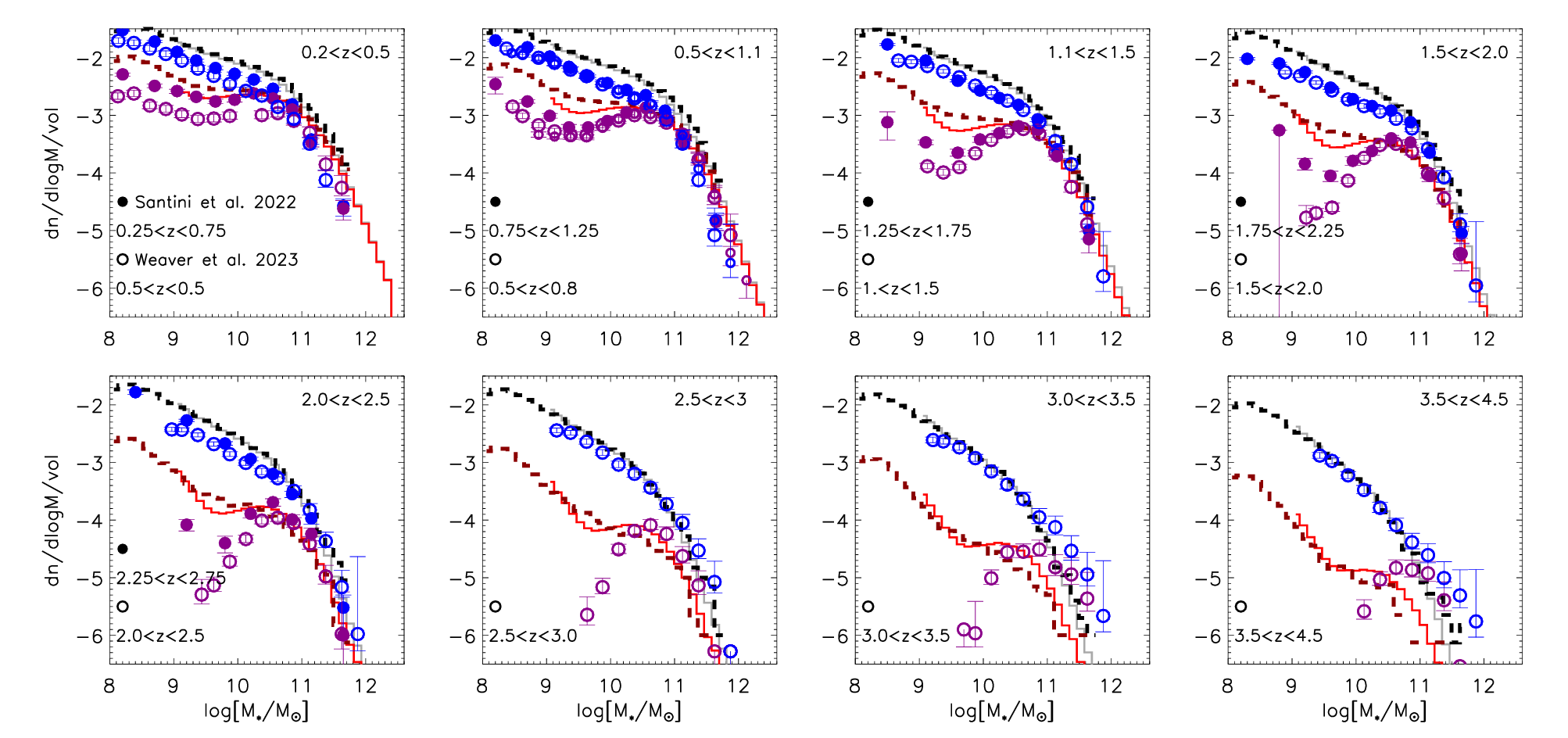}} 
\caption{Galaxy stellar mass function for all galaxies (gray/black) and for
  quiescent galaxies selected on the basis of their sSFR (red/dark red) as
  predicted from the \gaea model run on the Millennium Simulation (solid lines)
  and on the Millennium II (thick dashed lines). Symbols with error bars show
  different observational predictions as indicated in the legend. All model
  predictions shown in theis figure assume a stellar mass uncertainty of
  $0.25$~dex. The redshift bins used for the model have been optimized to match
  those employed by \citet{Weaver_etal_2023}, that are slightly offset from
  that adopted in \citet{Santini_etal_2022} as indicated in the legend.}
\label{fig:gmfMII}
\end{figure*}

Fig.~\ref{fig:gmfMII} shows the predicted galaxy stellar mass functions for all
(black/gray lines for model data and blue symbols for observations) and
quiescent (dark red/red lines and red symbols) galaxies. Solid lines correspond
to the Millennium Simulation while thick dashed lines show the corresponding
predictions for the Millennium II. All model predictions assume an
observational uncertainty on galaxy stellar masses of $0.25$~dex. Symbols with
error bars show different available observational estimates, as indicated in
the legend. Our \gaea model exhibits a very good convergence over the entire
galaxy mass range considered for the Millennium Simulation when considering all
model galaxies. The low-mass end of the quiescent galaxy stellar mass function
is somewhat more affected by resolution. However, a clear turn-over is still
evident at galaxy stellar masses between $\sim 10^{9}$ and $\sim 10^{10}\,{\rm
  M}_{\sun}$. Below this mass limit, our model appears to over-predict the
number densities measured for passive galaxies, and even more so when
considering predictions from the Millennium II simulation. As noted earlier,
this mass regime can be significantly affected by a different selection of
quiescent galaxies, with a colour-colour cut typically selecting less galaxies
than a cut based on sSFR. We defer a more detailed comparison to a future
study, where we will also push our model predictions to earlier cosmic
epochs. The last two panels of Fig.~\ref{fig:gmfMII} show that our new model
under-predicts the number densities of the most massive quiescent galaxies with
respect to observational estimates by \citet{Weaver_etal_2023}. We look into
this in more detail below.

\subsection{The cosmic stellar mass density evolution}

\begin{figure}
\centering
\resizebox{9cm}{!}{\includegraphics{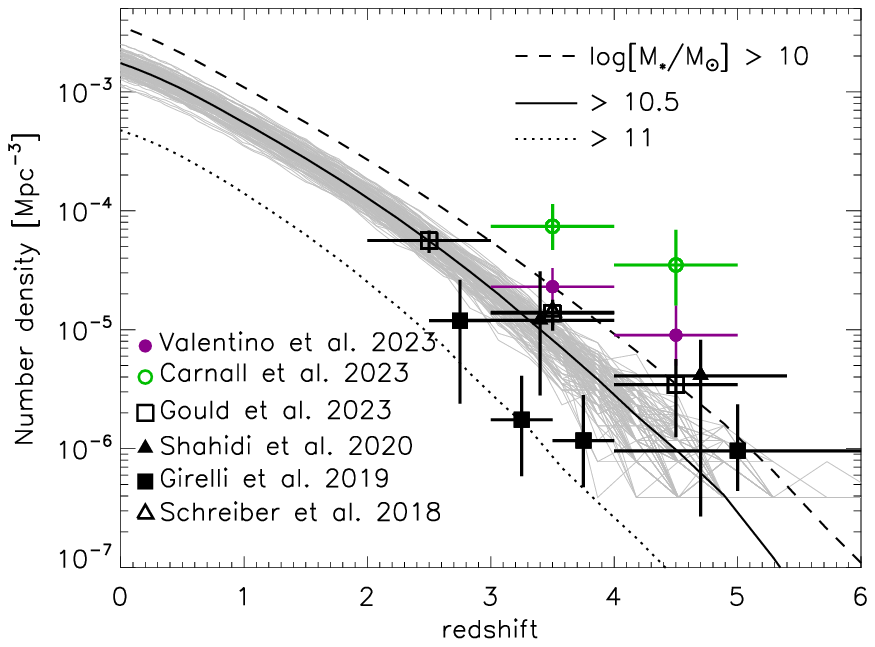}} 
\caption{Cumulative number density of massive quiescent galaxies as a function
  of redshift. Lines show model predictions corresponding to stellar mass cuts
  of $10^{10}$ (dashed), $10^{10.5}$ (solid - this is the closest to the
  observational measurements), and $10^{11}\,{\rm M}_{\sun}$ (dotted). Symbols
  with error bars show different observational estimates, as indicated in the
  legend. Measurements based on JWST are indicated in green
  \citep{Carnall_etal_2023} and magenta \citep{Valentino_etal_2023}. The gray
  lines show the cumulative number density evolution estimated in 125
  independent sub-volumes of 100~Mpc~${\rm h}^{-1}$ on a side. All model
  predictions assume an uncertainty on galaxy stellar masses of 0.25~dex.}
\label{fig:numdens_data}
\end{figure}

The number density of massive passive galaxies at early cosmic epochs has
gathered significant attention in recent years. In fact, the assembly of
massive galaxies in a relatively short cosmic time ($\sim 1.5-2$~billion years
when focusing on $z>3$) places strong constraints on galaxy formation
models. The topic has been subject to an intense debate over the years, and the
debate has been revived by recent studies based on JWST observations, showing
an excess of massive quiescent galaxies at $z>3$ with respect to previous
estimates \citep[e.g.][]{Carnall_etal_2023,Valentino_etal_2023}. Predictions
from our \gaea model are shown in Fig.~\ref{fig:numdens_data}, together with a
collection of recent pre- and post-JWST estimates. We show model predictions
corresponding to different stellar mass cuts, as indicated in the legend, and
assuming an uncertainty on stellar mass of 0.25~dex. The thick solid line is
the one closest to the cut adopted in all observational studies considered.
While our model is in quite good agreement with observational estimates based
on pre-JWST data, the figure confirms that it falls short of massive galaxies
at $z>3$ when considering the most recent observations. The gray lines show
model predictions obtained considering 125 independent sub-volumes of
100~Mpc~${\rm h}^{-1}$ on a side. These lines show that the expected cosmic
variance is large, which is not surprising considered that these are extreme
and rare galaxies. In fact, when taking the expected variance into account, our
model predictions are consistent with estimates by \citet{Valentino_etal_2023},
while still significantly below observational estimates by
\citet{Carnall_etal_2023}. As for most of the analysis presented in previous
sections, we have selected quiescent galaxies considering a time-dependent sSFR
cut. However, results do not vary significantly when considering a flat sSFR
cut corresponding to $10^{-10}\,{\rm yr}^{-1}$.

The comparison shown in Fig.~\ref{fig:numdens_data} should be considered with
caution: the data shown correspond to different selections of quiescent
galaxies, and rather different volumes. Estimates by \citet{Carnall_etal_2023}
are based on a total effective area of $\sim 30\,{\rm arcmin}^2$, while
\citet{Valentino_etal_2023} base their estimates on 11 JWST fields with
publicly available observations covering a total effective sky area of $\sim
145\, {\rm arcmin}^2$. The latter roughly correspond, at $3<z<5$, to a total
volume of $3.8\times10^6\,{\rm Mpc}^3$, approximately equal to the individual
sub-volumes considered to show the expected cosmic variance from our
model. \citet{Valentino_etal_2023} also find a significant field-to-field
variation in the estimated number densities, up to a factor 2-3. The presence
of over-densities in some of the areas studied could further decrease the
tension between our model predictions and the JWST estimates considered in
Fig.~\ref{fig:numdens_data}. As mentioned, the data shown in
Fig.~\ref{fig:numdens_data} are based on different colour-colour selections and
these could have a non-negigible impact at these redshift. Uncertanties in
photometric redshifts could also have an important impact on the estimated
number densities. Finally, it is also important to bear in mind that the
Millennium Simulation considered in this work has only 27 and 20 snapshots at
$z>3$ and $z>5$, respectively. It remains to be demonstrated that our model
results are stable on dark matter merger trees constructed on such a low number
of snapshots. We defer this to a future work.

\begin{figure}
\centering
\resizebox{9cm}{!}{\includegraphics{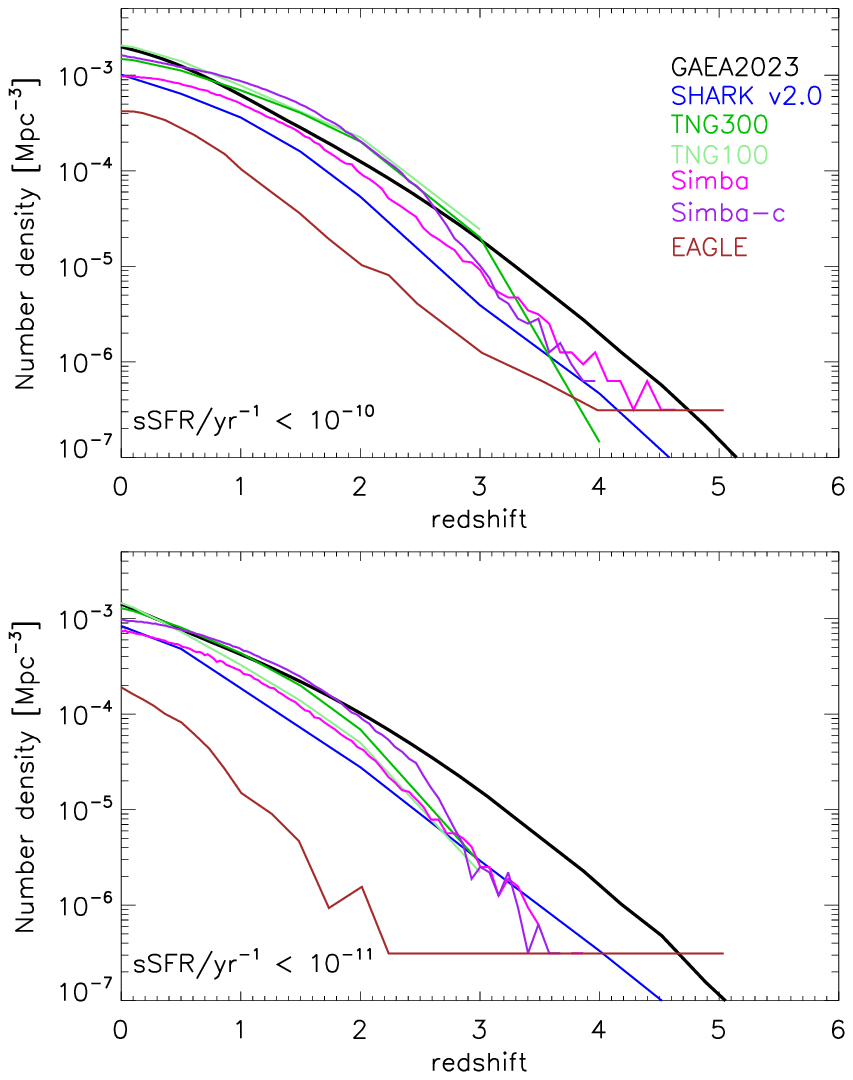}} 
\caption{Cumulative number density of massive quiescent galaxies as a function
  of redshift. Predictions from \gaea (shown by the black solid line in each
  panel) are compared with predictions from independent recently published
  models. The latter are shown as coloured lines, as indicated in the
  legend in the top panel. All model predictions correspond to galaxies more
  massive than $10^{10.5}\,{\rm M}_{\sun}$ and do not include any uncertainty
  on the properties considered (i.e. intrinsic predictions are shown). Top and
  bottom panels correspond to two different (flat) cuts in sSFR to select
  quiescent galaxies, as indicated in the bottom left of each panel.}
\label{fig:numdens_models}
\end{figure}

In order to put our model predictions in a context, we compare them to those
from independent recently published theoretical models in
Fig.~\ref{fig:numdens_models}. Specifically, we compare our model predictions
with results from the latest version of the SHARK semi-analytic model (in blue,
\citealt{Lagos_etal_2023}), from TNG100 and TNG300 (light and dark green
respectively, \citealt{Nelson_etal_2019}), two different versions of Simba
(magenta and purple, \citealt{Dave_etal_2019} and \citealt{Hough_etal_2023}),
and EAGLE (brown, \citealt{McAlpine_etal_2016} and references therein). We show
only number densities corresponding to quiescent galaxies more massive than
$10^{10.5}\,{\rm M}_{\sun}$. We have considered, in all cases, two flat cuts in
sSFR (top and bottom panels, as indicated in the legend). Predictions shown in
this figure do not assume any uncertainty in the physical properties used
(i.e. intrinsic values of SFR and stellar mass are used). The models considered
cover a range of number densities, at all redshifts. As discussed above, this
is partially due to different simulated volumes. However, based on results
shown in Fig.~\ref{fig:numdens_data}, volume differences cannot be the main
drivers of the differences between the models shown in
Fig.~\ref{fig:numdens_models}: TNG100 and TNG300, based on boxes of 100 and
300~Mpc respectively, predict number density evolutions that are much closer to
each other than to predictions from any other simulation considered. The same
is true for Simba and Simba-c (corresponding to boxes of size $\sim 150$ and
$\sim 74$~Mpc on a side, respectively). Predictions from EAGLE correspond to a
simulated volume comparable to that of TNG100 and Simba, while predictions from
SHARK are based on a box of $\sim 300$~Mpc, comparable to that of TNG300. While
this is smaller than the volume of the Millennium Simulation, it is larger than
the sub-boxes considered in Fig.~\ref{fig:numdens_data}. In summary, we believe
that the differences shown in Fig.~\ref{fig:numdens_models} are largely driven
by a different treatment of the baryonic processes.

At $z=0$, the EAGLE simulation predicts the lowest number density of quiescent
galaxies. \gaea and the TNG simulations predict the largest values of number
densities - these are larger than those predicted by EAGLE by about one order
of magnitude. The number densities of massive quiescent galaxies decrease
rapidly with increasing redshift, for all models. There are no massive
quiescent galaxies above $z\sim 3.7$ in TNG100, and above $z\sim 4$ in
TNG300. The decrease is smoother for the EAGLE simulation that however also
predicts number densities of massive quiescent galaxies of only $\sim
10^{-6}\,{\rm Mpc}^{-3}$ at $z=3$ when considering a sSFR cut $<10^{-10}\,{\rm
  yr}^{-1}$. Among all models considered, \gaea predicts the largest number
densities at $z>3$.

\section{Summary and conclusions}

In this paper, we present an updated version of our GAlaxy Evolution and
Assembly (GAEA) theoretical model of galaxy formation
\citep{DeLucia_etal_2014,Hirschmann_etal_2016}. In particular, the latest
version of the model ({\sc GAEA2023}) now combines for the first time (i) an
updated treatment of AGN feedback that includes an improved modelling of cold
gas accretion on super-massive black holes and an explicit implementation of
AGN-driven outflows \citep{Fontanot_etal_2020}; and (ii) an improved modelling of both
cold and hot gas stripping from satellite galaxies \citep{Xie_etal_2020}. We
have shown that our latest model version predicts specific star formation rate
(sSFR) distributions, for galaxies of different stellar mass, that are in
remarkable agreement with observational measurements in the local Universe.

Comparing with previous versions of our model, we can characterize the physical
drivers of the improved agreement with data: the updated treatment of
satellite galaxies, featuring a non instantaneous gas stripping, leads to
relatively long quenching time-scales for low-mass galaxies and a sSFR
distribution with a single peak that coincides with the value observed in the
local Universe. The inclusion of quasar winds leads to a rapid suppression of
the star formation, that manifests itself as a clear bimodality in the sSFR
distributions for galaxies with stellar masses between $10^{9.5}$ to
$10^{11}\,{\rm M}_{\sun}$, and no excess of star formation for more massive
galaxies. We find that the improved treatment of the density threshold for star
formation, resulting from our explicit modelling of the partition between atomic
and molecular hydrogen \citep{Xie_etal_2017,Xie_etal_2020}, does play an
important role for the most massive galaxies by avoiding an extremely bursty
and unrealistic behaviour for their star formation histories.

Predictions from our updated model version are also in quite nice agreement
with observational measurements of the quenched fractions up to $z\sim
3-4$. This again improves significantly over our previous models that were
predicting either too-low fractions of quiescent galaxies at $z>1$ \citep[][see
  also \citealt{DeLucia_etal_2019}]{Xie_etal_2020}, or an important excess of
quenched galaxies below stellar masses $\sim 10^{10}\,{\rm M}_{\sun}$
\citep{Fontanot_etal_2020}. We show that all models discussed in this work
predict a turn-over of the number densities of quiescent galaxies at
low-masses, with the exact location of this turn-over depending on the
model considered and being closest to observational estimates for our latest
model version. This is a robust prediction of our models, and it is not
affected by limited resolution. Deeper observations, that can attempt to
  constrain the location of this turnover at intermediate to high redshift,
can therefore provide important constraints on the physical processes
regulating the number densities of low-mass quiescent galaxies, and in
particular on the relative importance of stellar feedback and environmental
processes in suppressing star formation in these systems.

As discussed in recent work \citep[e.g.][]{Kukstas_etal_2023}, state-of-the-art
theoretical models struggle to reproduce the observed quenching of galaxies in
environments that are satellites dominated, i.e. in galaxy clusters. Our
updated model is in remarkable agreement with observational estimates at $z\sim
1$, when the (naturally large) expected halo-to-halo variation is taken into
account. However, a comparison with available observational measurements of the
quiescent galaxy mass function suggests that the model still slightly
over-predicts the fractions of quenched galaxies at the lowest stellar masses
sampled, and under-predicts the quenched number densities around the
knee. I.e. the overall shape of the galaxy stellar mass function for quiescent
and star forming galaxies is very similar at intermediate galaxy stellar
masses, in contrast with available data. This regime is sensitive to both
stellar and AGN feedback, so accurate measurements for larger samples of
clusters might provide very useful constraints on these physical processes.

The bulk of the analysis just summarized is based on an evolving cut in sSFR to
select quiescent galaxies, while most observational estimates at $z>0$ are
based on some colour-colour selection. Using a sub-volume of the simulation, we
show that a $UVJ$ colour-colour selection would not have a significant impact
on the discussed agreement between model predictions and observational
estimates up to $z\sim 3$. At larger redshift, a classical $UVJ$ diagram fails
to identify increasingly larger fractions of galaxies that are intrinsically
quiescent (i.e. have very low levels of sSFR) in \gaea. We also demonstrate
that a different selection might impact significantly the number densities of
quiescent galaxies at stellar masses $<10^{10}\,{\rm M}_{\sun}$. In future
studies, we plan to carry out a more detailed comparison with observational
data by mimicking the adopted colour-colour selections and extending the
analysis to higher redshifts than considered in the present work.

Our latest model predicts number densities of massive quiescent galaxies at
$z>3$ that are larger than those predicted by a number of recently published
state-of-the-art models. Yet, our model predictions appear to be below JWST
observational measurements \citep{Carnall_etal_2023,Valentino_etal_2023}. We
show that the expected cosmic variance is very large, and it can easily
accomodate at least some of the recently published measurements. More careful
comparisons, based on the same criteria to select quiescent galaxies, are
needed. Caution is also needed when interpreting these results due to the fact
that the simulations used in this work do not have a fine sampling of the dark
matter merger trees at these early cosmic epochs. In future work, we will
re-address this issue by taking advantage of a larger simulation with much
finer time sampling at high redshift.

\begin{acknowledgements}
  We acknowledge the use of INAF-OATs computational resources within the
  framework of the CHIPP project \citep{Taffoni_etal_2020}.  We thank Claudia
  Lagos, Romeel Dav\'e, and Jakub Szpila for making results from SHARK v2.0,
  Simba, and Simba-c available in electronic format. We acknowledge stimulating
  discussions with Adam Muzzin, Veronica Strazzullo, and John Weaver. MH
  acknowledges funding from the Swiss National Science Foundation (SNSF) via a
  PRIMA grant PR00P2-193577 ‘From cosmic dawn to high noon: the role of BHs for
  young galaxies’. We acknowledge the anonymous referee and Veronica
  Strazzullo, for careful reading of our manuscript and many useful suggestions
  that helped improving the clarity of our presentation.
\end{acknowledgements}

%
\bibliographystyle{aa} 
\bibliography{mergmodel} 
%
\begin{appendix} 
\label{app:app1}
\section{Model calibration}
\end{appendix}

\begin{figure*}
\centering
\resizebox{16cm}{!}{\includegraphics{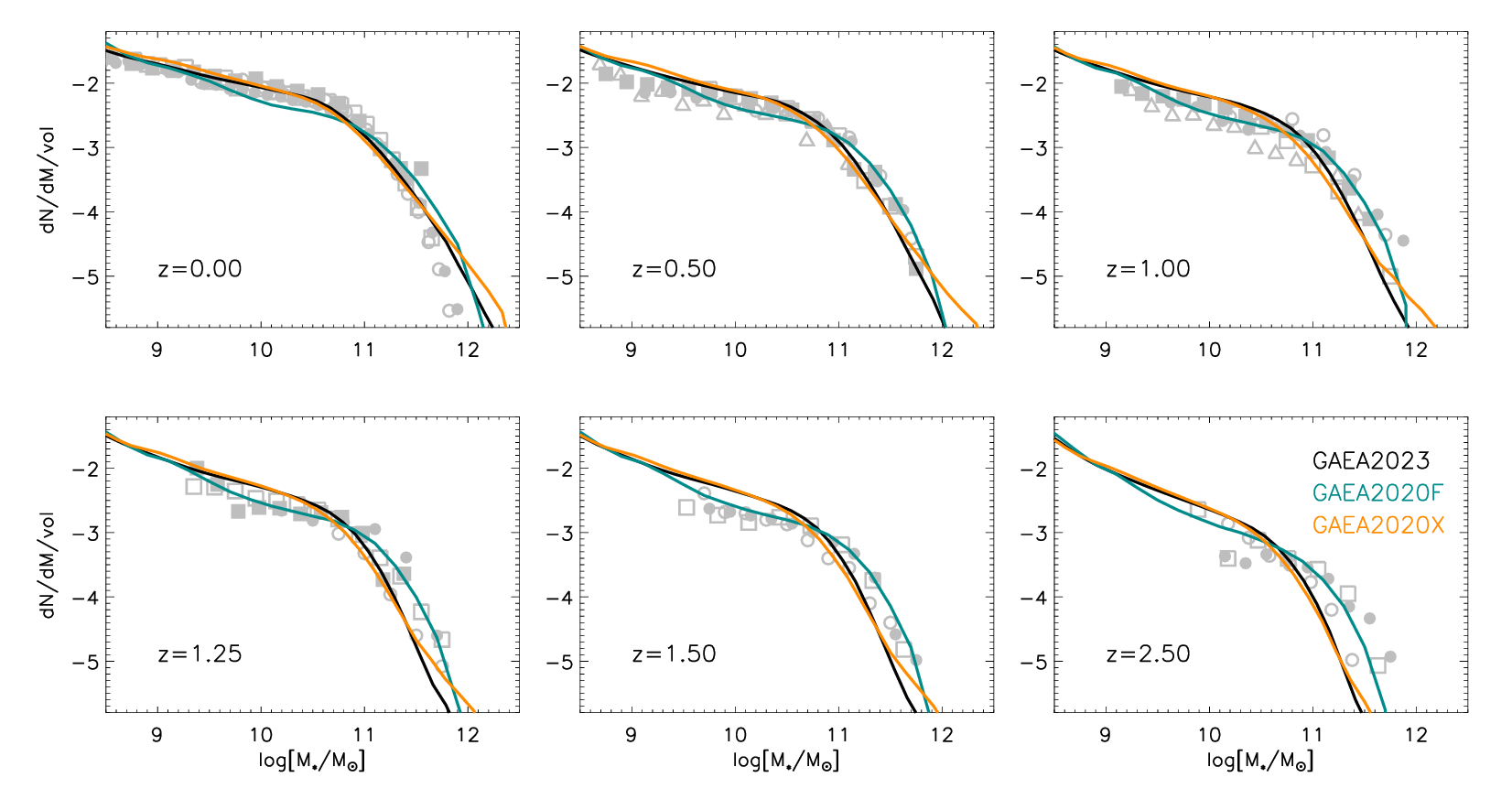}} 
\caption{The galaxy stellar mass functions at different redshifts. Solid lines show predictions all three models considered in this paper, while symbols show the same compilation of observational data considered in \citet{Hirschmann_etal_2016}.}
\label{fig:gmf}
\end{figure*}

\begin{figure*}
\centering
\resizebox{16cm}{!}{\includegraphics{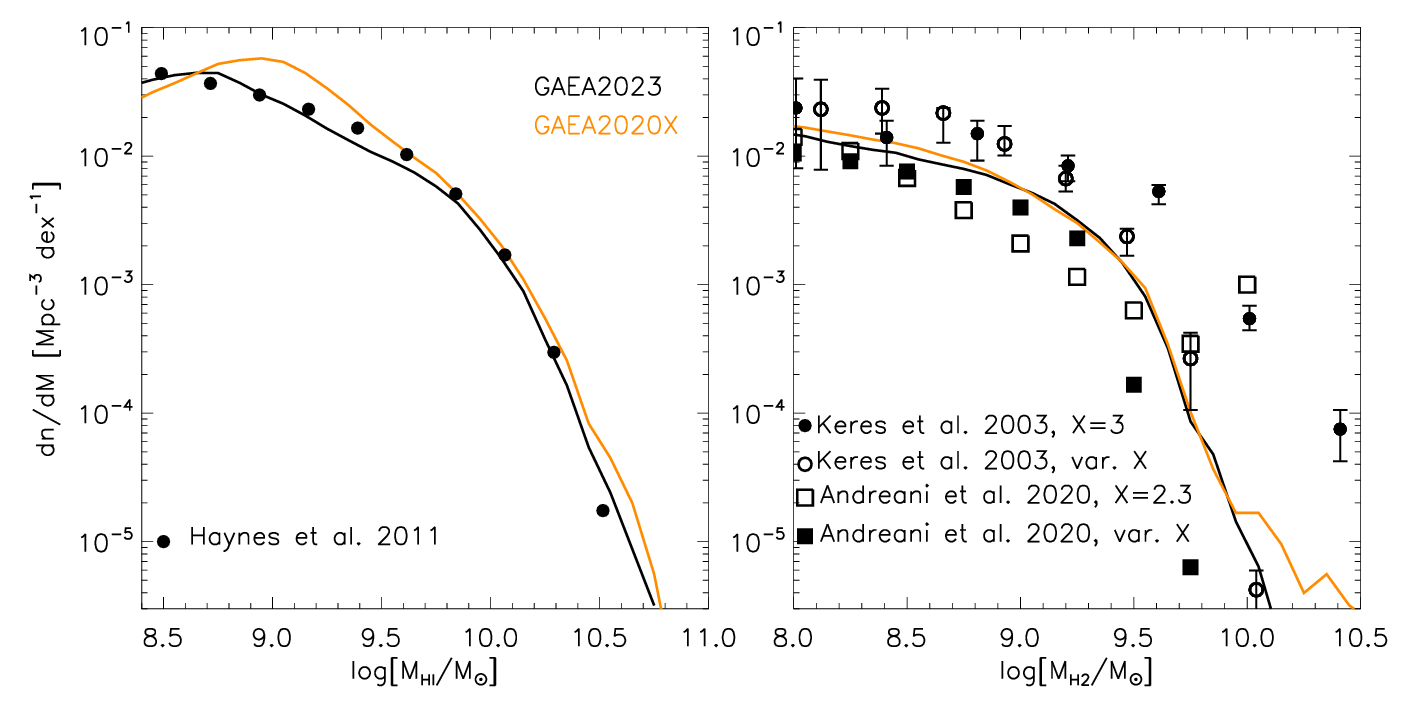}} 
\caption{The HI (left panel) and H$_2$ (right panel) mass function at $z=0$,
  compared with observational estimates. The black solid line corresponds to
  our \gaea model, while the dark orange line shows model predictions from
  \citet{Xie_etal_2020}.}
\label{fig:HIH2MF}
\end{figure*}

\begin{figure*}
\centering
\resizebox{16cm}{!}{\includegraphics{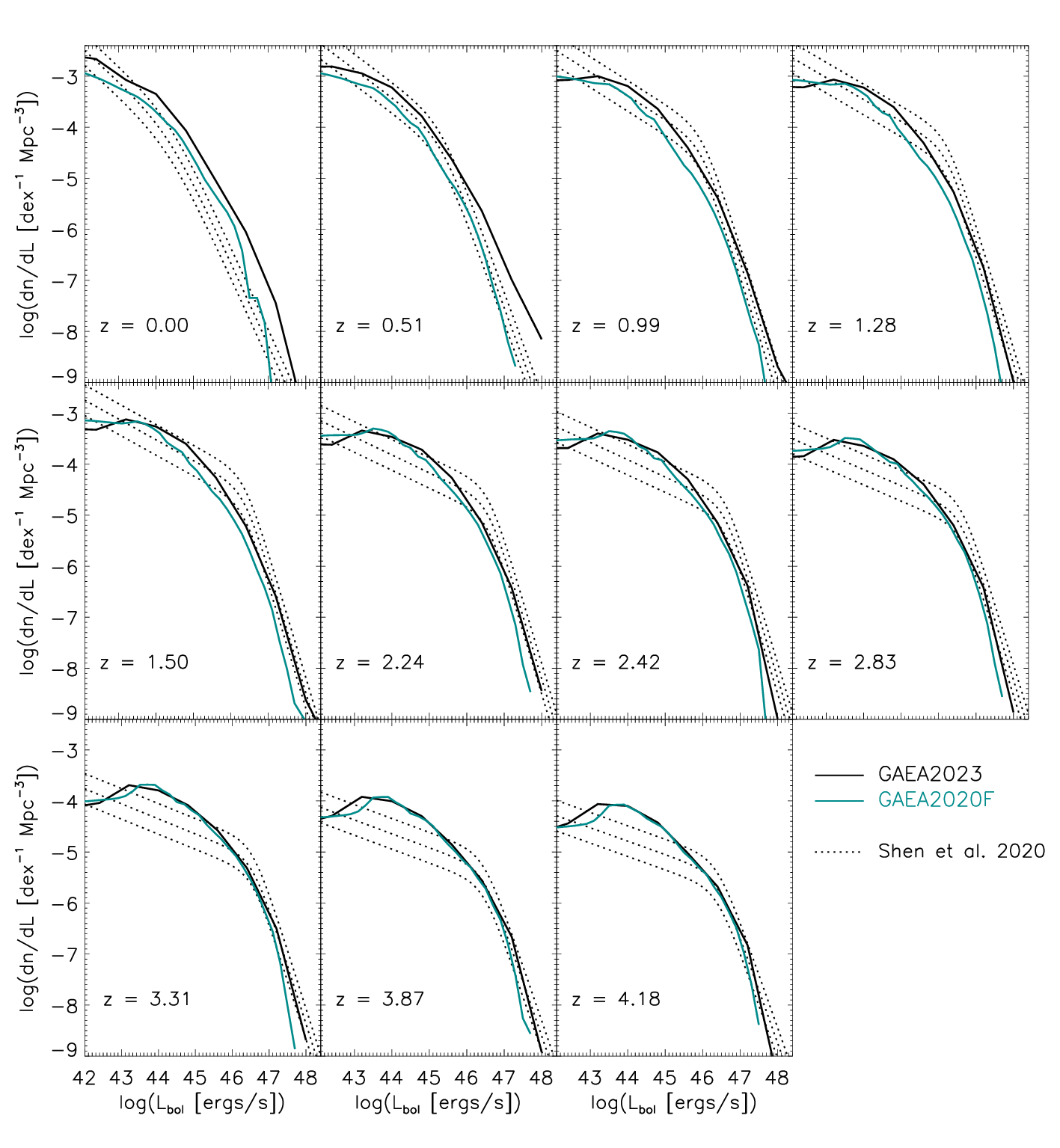}} 
\caption{The AGN bolometric luminosity function at different redshifts. Solid
  lines show predictions from the \gaea and \gaeaf models, while the solid
  lines correspond to observational estimates by \citet{Shen_etal_2020}
  assuming an uncertainty of +/-0.3~dex..}
\label{fig:qsolf}
\end{figure*}

As mentioned in Section~\ref{sec:simsam}, the primary observables that have
been considered to recalibrate our model are: the observed galaxy stellar mass
function and its evolution up to $z\sim 3$, the HI and H$_2$ mass functions in
the local Universe, the evolution of the AGN luminosity function up to $z\sim
4$. For completeness, we show in this Appendix the main calibration plots,
including all three models that have been presented in this
paper.

Fig.~\ref{fig:gmf} shows the predicted galaxy stellar mass function at
different redshifts, from all three models considered in this paper, compared
with a compilation of observational data as in \citet{Hirschmann_etal_2016}.
Fig.~\ref{fig:HIH2MF} shows the HI galaxy mass function in the left panel,
compared with observational estimates by \citet{Haynes_etal_2011}, and the
H$_2$ galaxy mass function, compared with observational estimates by
\citet{Keres_etal_2003} and \citet{Andreani_etal_2020}. Finally,
Fig.~\ref{fig:qsolf} shows the predicted AGN luminosity function from the \gaea
and \gaeaf models, compared with observational estimates by
\citet[][we consider an uncertainty of +/-0.3~dex]{Shen_etal_2020}.

\end{document}